\documentclass[a4paper,11pt]{article}
\pdfoutput=1 % if your are submitting a pdflatex (i.e. if you have images in pdf, png or jpg format)
\usepackage{jheppub} % for details on the use of the package, please see the JHEP-author-manual
\usepackage[T1]{fontenc} % if needed
\usepackage{natbib}
\usepackage{tensor}
\usepackage{soul,xcolor}
\usepackage{diagbox}
\def\ba{\begin{eqnarray}}
\def\ea{\end{eqnarray}}
\newcommand{\rmd}{{\rm d}}

\newcommand{\F}{\mathcal{F}}
\newcommand{\beq}{\begin{equation}}
\newcommand{\eeq}{\end{equation}}
\newcommand{\bal}{\begin{aligned}}
\newcommand{\eal}{\end{aligned}}

\title{\boldmath Cubic interactions for massless and partially massless spin-1 and spin-2 fields}

\author[a]{Nicolas Boulanger,}
\author[b]{Sebastian Garcia-Saenz,}
\author[b]{Songsong Pan,}
\author[a]{Lucas Traina}

\affiliation[a]{
	Physique de l'Univers, Champs et Gravitation, Universit\'e de Mons – UMONS, \\
	Place du Parc 20, 7000 Mons, Belgium
}

\affiliation[b]{
	Department of Physics, Southern University of Science and Technology, \\
	Shenzhen 518055, China
}

\emailAdd{nicolas.boulanger@umons.ac.be}
\emailAdd{sgarciasaenz@sustech.edu.cn}
\emailAdd{12232930@mail.sustech.edu.cn}
\emailAdd{lucas.traina@alumni.umons.ac.be}

\abstract{
	We perform a complete classification of the consistent two-derivative cubic couplings for a system containing an arbitrary number of massless spin-1, massless spin-2, and partially massless (PM) spin-2 fields in $D$-dimensional (anti-)de Sitter space. In addition to previously known results, we find a unique candidate mixing between spin-1 and PM spin-2 fields. We derive all the quadratic constraints on the structure constants of the theory, allowing for relative ``wrong-sign'' kinetic terms for any of the fields. In the particular case when the kinetic terms in each sector have no relative signs, we find that the unique consistent non-trivial theory is given by multiple independent copies of conformal gravity coupled to a Yang-Mills sector in $D=4$. Our results strengthen the well-known no-go theorems on the absence of mutual interactions for massless and PM spin-2 fields.
}

\begin{document} 
\maketitle
\flushbottom

\section{Introduction and summary} \label{sec:intro}

De Sitter (dS) and anti-de Sitter (AdS) spaces are natural and well-motivated 
arenas to study theories of higher-spin gauge fields. Vasiliev theory 
\cite{Vasiliev:1990en,Vasiliev:1992av,Vasiliev:2003ev} provides a striking example 
of how the obstructions one encounters in attempts to construct higher-spin theories 
in flat space can be evaded in AdS (see \cite{Vasiliev:1999ba,Bekaert:2010hw} for 
reviews). 
On the other hand, observational evidence indicates that our universe was very 
approximately a dS space during the pre-hot Big Bang  
epoch and will evolve towards a dS space in the future. 
The high energy scales that could potentially be probed by cosmological experiments 
therefore motivate a good understanding of the dynamics of higher-spin particles 
in dS. Although this mostly concerns massive fields rather than gauge fields, one may 
envisage a Goldstone equivalence regime in which the massive theory is described by a 
tower of higher-spin gauge fields, or speculate about higher spin analogs 
of the Brout-Englert-Higgs or dynamical symmetry breaking
mechanisms --- see \cite{Boulanger:2006tg} and \cite{Girardello:2002pp}, respectively.
Additionally, and more to the point of this paper, (A)dS space allows for 
the existence of exotic ``massive'' particles described by gauge fields, 
that one calls partially massless (PM) fields \cite{Deser:1983mm,Deser:2001pe,Deser:2001us,Zinoviev:2001dt}. 
While bosonic PM fields are non-unitary in AdS, in dS they precisely saturate 
the Higuchi bound \cite{Higuchi:1986py,Deser:2001wx}, meaning that they would 
have behaved as light fields during a putative phase of cosmic inflation, with 
interesting observational imprints \cite{Baumann:2017jvh}.

Similarly to their Fronsdal cousins \cite{Fronsdal:1978rb}, interactions of 
higher spin PM fields are highly constrained by their gauge structure, 
and in fact no fully satisfactory examples of interacting higher-spin 
theories with PM fields are known (see however \cite{Brust:2016xif,Brust:2016zns} 
for an intriguing proposal of a Vasiliev-like PM theory, and \cite{Grigoriev:2020lzu} 
for a proposal in three dimensions). 
While a few interesting studies have tackled the problem in the case of 
PM particles with spin $s>2$ \cite{Joung:2012hz,Joung:2012rv,Sleight:2021iix}, 
the most detailed analyses have been focused on PM fields of spin $s=2$ \cite{deRham:2013wv,Hinterbichler:2014xga,Boulanger:2018shp,Hinterbichler:2015nua,Garcia-Saenz:2014cwa,Bonifacio:2016blz,Bernard:2017tcg,Boulanger:2019zic}. 
One reason is that this is of course the most technically tractable case, 
but there is also the more physical motivation that PM spin-2 fields may have some 
connection with theories of massive gravity \cite{deRham:2012kf,Deser:2013uy,Hassan:2012gz,Hassan:2012rq,Hassan:2013pca,Hassan:2015tba}. 
Cubic couplings involving PM gravitons have in particular been subject of several 
studies \cite{Zinoviev:2014zka,Garcia-Saenz:2015mqi,Goon:2018fyu,Boulanger:2019zic}. 
It is by now well established that three-point vertices for PM spin-2 particles are forbidden, at least if one assumes up to two-derivative interactions 
\textit{and} the absence of ``ghost-like'' fields, i.e.\ fields with wrong-sign 
kinetic term. 
This no-go result has been extended 
in Ref.\ \cite{Joung:2014aba} to also include 
the gravitational coupling with a massless spin-2 field.

Although the assumption of unitarity or, more specifically in the present context, 
absence of negative-norm states may seem essential, it is important to bear in mind that 
a complete non-linear theory that couples a PM graviton 
with a massless graviton in fact exists: 
conformal gravity \cite{Maldacena:2011mk,Deser:2012euu,Deser:2012qg}. 
More in detail, conformal gravity in $D=4$ dimensions, upon linearizing the 
theory about an (A)dS background, yields a massless and a PM spin-2 modes, 
one of which is necessarily ghostly.\footnote{This may be generalized to any 
higher even dimension $D$: the spectrum contains $D/2$ spin-2 modes, 
one of which is massless, one is PM, and the rest are massive \cite{Joung:2019wwf}.} 
Although the physical viability of conformal gravity is perhaps questionable due to this 
latter property, it is nevertheless a model of obvious theoretical interest; see 
\cite{Mannheim:2011ds,Fradkin:1985am} for reviews, including its supersymmetric 
extension. It is therefore a tantalizing possibility that other non-unitary theories that 
include PM fields may exist. This is further motivated by the results of Refs.\ \cite{Garcia-Saenz:2018wnw,Bittermann:2020xkl,Buchbinder:2019olk} on the classification of supersymmetric multiplets that include PM fields in four dimensions: 
PM supersymmetric representations are always non-unitary\footnote{Note that unitarity 
could be restored in dS background, see 
\cite{Letsios:2022tsq,Letsios:2023qzq,Letsios:2023tuc} for more explanations.}; 
in particular, the simplest multiplet contains a PM spin-2 and a massless spin-1 field in its 
bosonic sector, and these must be relatively ghostly.
The fermionic sector contains a massive 
and a massless spin-3/2 fields. Cubic couplings for PM spin-2 and massive spin-3/2 particles 
have been recently classified in \cite{Boulanger:2023lgd} (see also \cite{Zinoviev:2018eok} 
for earlier work). Interestingly, the system containing relatively ghostly PM spin-2 and a 
massless spin-1 fields has been shown to be conformally invariant in $D=4$ dimensions 
\cite{Farnsworth:2024iwc}.

In line with these considerations we mention two works that have studied the consequences 
of relaxing the assumption of positivity of kinetic terms in the construction of 
two-derivative cubic vertices involving PM spin-2 fields. 
The first work \cite{Joung:2019wwf} 
considered the inclusion of massive and massless spin-2 fields, finding the cubic vertex 
of even-dimensional conformal gravity as a consistent solution, among other candidates. 
The second work \cite{Boulanger:2019zic} restricted its attention to only PM fields, 
showing that there exists a cubic vertex which is unique and  
consistent at the full non-linear level. Once again, these results necessitate 
non-positive-definite kinetic (or ``internal'') matrices, but are otherwise 
consistent from the point of view of the gauge structure.
Remarkably, the model of \cite{Boulanger:2019zic} is the only consistent, 
interacting theory for a multiplet of PM spin-2 fields, although non-unitary 
at the classical level.

In the analysis of Ref.\ \cite{Joung:2019wwf} the additional spin-2 fields were introduced in 
an ad-hoc fashion so as to ensure consistency of the PM gravitational coupling. This motivates 
us to revisit the problem of classifying interactions for massless and PM spin-2 fields, 
assuming a more general starting point in which the number of fields and the signature of their 
internal metrics are arbitrary. Furthermore, again prompted by the results on supersymmetric 
representations as well as by the related findings of Ref.\ 
\cite{Boulanger:2023lgd,Farnsworth:2024iwc}, we also consider the inclusion 
of massless spin-1 fields in the spectrum. Let us summarize our main results:
\begin{itemize}
\item We provide a complete classification of the consistent first-order deformations, in the sense of the Noether procedure, of the free theory describing an arbitrary collection of massless spin-1, massless spin-2 and PM spin-2 fields in rigid $D$-dimensional (A)dS space. Our classification is completely general in regards to deformations of the gauge algebra, but is restricted to at most four derivatives in the deformations of the gauge symmetries and to at most two derivatives in the cubic vertices. A further restriction is that we focus on parity-even deformations.

Our results confirm the classification of Ref.\ \cite{Joung:2019wwf} for massless and PM spin-2 fields. We also find a unique candidate vertex mixing massless spin-1 and PM spin-2 particles. This vertex is of the Chapline-Manton type, i.e.\ it is Abelian yet induces a non-linear gauge transformation of the spin-1 fields. Moreover, it only exists in $D=4$ dimensions.

\item We analyze the consistency of the candidate deformations at the next order 
in perturbation, starting with the Jacobi identities for the candidate gauge algebras. 
We derive all the quadratic constraints on the structure constants obtained in the 
previous step. The results are valid for 
any choice of the internal metrics that define the kinetic terms of the fields, thus 
accommodating any choice of healthy/ghostly field content.

We consider the most general solution of the constraints under the assumption that each field sector (spin-1, massless spin-2 and PM spin-2) contains no relative `healthy/ ghostly' signs in the kinetic terms, although distinct sectors may do so. We find the answer to be given by multiple, independent copies of $D=4$ conformal gravity minimally coupled with a Yang-Mills (or possibly Abelian) spin-1 sector.

This outcome again confirms the conclusions of Ref.\ \cite{Joung:2019wwf} with 
regards to the spin-2 fields, although with a more general starting point. Our 
results also imply that the doubled spectra model identified in that reference in 
fact corresponds to two non-interacting copies of conformal gravity and is therefore 
not a new theory.\footnote{Note that the consistent deformations of 
(multi-)conformal (or Weyl) gravity were investigated in \cite{Boulanger:2001he}.}
Moreover, we can rule out the consistency of the 
non-geometric vertex identified in \cite{Joung:2019wwf}\footnote{This vertex
actually is the contraction of the massless spin-2 field $h_{\mu\nu}$ with 
the PM spin-2 current $J^{\mu\nu}$ first identified in \cite{Zinoviev:2006im}
in the context of PM spin-2 self interactions.}, 
at least under the aforementioned assumptions. An additional corollary is that 
distinct massless graviton species cannot mutually interact through the exchange of 
massless spin-1 or PM spin-2 particles, thus further generalizing the well-known 
no-go theorem of Ref.\ \cite{Boulanger:2000rq}.

Finally, we discuss some solutions to the quadratic constraints in the more general 
set-up with non-sign-definite internal metrics. Although our analysis is not 
exhaustive, we are able to exhibit particular solutions for which all the candidate 
vertices remain consistent.
\end{itemize}

Our analysis makes use of the 
Becchi-Rouet-Stora-Tyutin-Batalin-Vilkovisky
(BRST-BV) \cite{Becchi:1975nq,Tyutin:1975qk,Batalin:1981jr,Batalin:1983ggl} 
reformulation of the Noether procedure along the cohomological lines of  
\cite{Barnich:1993vg,Henneaux:1997bm}. 
This is an ideally well-suited technique to deal with the usual ambiguities 
related to field and gauge parameter redefinitions, essentially recasting 
and generalizing the procedure of \cite{Berends:1984rq}
in the form of a well-defined cohomological problem in the presence of antifields.  
Our results are thus guaranteed to be general and unambiguous within the stated 
assumptions. We briefly review the method and formulate its application to our 
system in Sec.\ \ref{sec:brst-bv}. In Sec.\ \ref{sec:firstorder} we present the 
results of the first-order deformation analysis, continuing in 
Sec.\ \ref{sec:secondorder} with the study of the second-order consistency 
and derivation of quadratic constraints. Finally in Sec.\ \ref{sec:analysis} we 
investigate the resolution of the quadratic constraints and 
briefly summarize our finding 
in Sec.\ \ref{sec:conclusions}.

%%%%%%%%%%%%%%%%%%%%%%%%%%%%
%%%%%%%%%%%%%%%%%%%%%%%%%%%%
\section{BRST-BV formulation} \label{sec:brst-bv}

The spectrum of fields considered in this paper consists of 
an arbitrary collection of $n_g$ tensor fields $h^I_{\mu\nu}$ 
describing massless spin-2 fields, $n_{\rm PM}$ tensor fields 
$k^{\Delta}_{\mu\nu}$ describing PM spin-2 fields, and $n_v$ vector fields 
$A^a_{\mu}$ describing massless spin-1 fields. 
The tensors $h^I_{\mu\nu}$ and $k^{\Delta}_{\mu\nu}$ are symmetric 
in their lower indices.
See Table \ref{tab:field content} for a summary of our notations.

\begin{table}[ht]
\centering
\begin{tabular}{|l|c|c|c|}
\hline
                & Field variable        & Curvature                           & Indices                                   \\ \hline
Massless spin-2 & $h^I_{\mu\nu}$        & $K^I_{\mu\nu\rho\sigma}$            & $I,J,\ldots\in\{1,\ldots,n_g\}$           \\ \hline
PM spin-2       & $k^{\Delta}_{\mu\nu}$ & $\F^{\Delta}_{\mu\nu\rho}$ & $\Delta,\Sigma,\ldots\in\{1,\ldots,n_{\rm PM}\}$ \\ \hline
Massless spin-1 & $A^a_{\mu}$           & $F^a_{\mu\nu}$                      & $a,b,\ldots\in\{1,\ldots,n_v\}$           \\ \hline
\end{tabular}
\caption{Field content and notations for the system considered in this paper.}
\label{tab:field content}
\end{table}

Our starting point is the non-interacting action for the free propagation of the fields on a rigid $D$-dimensional (A)dS space,
\begin{equation}
\begin{aligned}
S_0\left[h_{\mu \nu}^I, k_{\mu \nu}^{\Delta}, A_\mu^a\right]&=\int \rmd^D x\sqrt{-g}\bigg[\mathfrak{g}_{IJ}  \bigg(-\frac{1}{2} \nabla^\rho h^{I\mu \nu} \nabla_\rho h_{\mu \nu}^J+\nabla_\rho h^{I \mu \nu} \nabla_\mu h_\nu^{J \rho}-\nabla_\mu h^I \nabla_\nu h^{J \mu \nu} \\
&\quad  \left.+\frac{1}{2} \nabla^\mu h^I \nabla_\mu h^J-\left(\frac{D-1}{\sigma L^2}\right) h^{I \mu \nu} h_{\mu \nu}^J+\frac{1}{2}\left(\frac{D-1}{\sigma L^2}\right) h^I h^J\right) \\
&\quad + \mathfrak{g}_{\Delta \Omega}\left(-\frac{1}{4} \F^{\Delta \lambda \mu \nu} \F_{\lambda \mu \nu}^{\Omega}+\frac{1}{2} \F^{\Delta \lambda} \F_\lambda^{\Omega}\right)-\frac{1}{4}\,\mathfrak{g}_{a b}\,F^{a \mu \nu} F_{\mu \nu}^b \bigg] \,.
\end{aligned}
\end{equation}
Here $L$ is the radius of the (A)dS space and $\sigma$ is a sign, $+1$ for AdS 
and $-1$ for dS. Spacetime (greek) indices are moved with the (A)dS metric 
$g_{\mu\nu}$ and $\nabla$ is the corresponding metric-compatible covariant 
derivative. We write $h^I:=g^{\mu\nu}h^I_{\mu\nu}$ and 
$k^{\Delta}:=g^{\mu\nu}k^\Delta_{\mu\nu}$. 
As explained in the Introduction, we have allowed for arbitrary (constant) 
`internal' field space metrics $\mathfrak{g}_{IJ}$, $\mathfrak{g}_{\Delta \Omega}$ 
and $\mathfrak{g}_{a b}$. Through trivial field redefinitions these may be brought to 
the form $\operatorname{diag}(+,\ldots,+,-,\ldots,-)$, and a `unitary' theory 
corresponds to the case with positive definite metrics.\footnote{We say `unitary' 
with some abuse of terminology, since actually PM spin-2 fields are anyway non-
unitary in AdS, irrespective of the signature of 
$\mathfrak{g}_{\Delta \Omega}$ \cite{Deser:2001pe}.} 
All the internal `color' indices are raised and lowered with these metrics.
The above action features the tensors 
$\F_{\lambda \mu \nu}^{\Delta}:=2\, \nabla_{[\lambda} k_{\mu] \nu}^{\Delta} \,$ 
and $\F^{\Delta}_{\lambda}:=g^{\mu\nu}\F_{\lambda \mu \nu}^{\Delta}\,$,
as well as $F^a_{\mu\nu}:=2\,\nabla_{[\mu}A^a_{\nu]} \,$.
The action $S_0$ is invariant under the following gauge transformations
\begin{equation}
\begin{aligned}
\delta_0 h_{\mu \nu}^I & =2 \,\nabla_{(\mu}\epsilon_{\nu)}^I \,, \\
\delta_0 k_{\mu \nu}^{\Delta} & =\nabla_\mu \nabla_\nu \epsilon^{\Delta}-\frac{\sigma}{L^2}g_{\mu \nu} \epsilon^{\Delta} \,, \\
\delta_0 A_\mu^a & =\nabla_\mu \epsilon^a \,.
\end{aligned}
\end{equation}
The gauge parameters $\epsilon^I_{\mu}$, $\epsilon^\Delta$ and $\epsilon^a$ are 
arbitrary functions and the gauge symmetries are irreducible, i.e.\,, 
there is no gauge-for-gauge transformations.
The following linearized curvatures or field strengths are invariant 
under the above gauge transformations:
\begin{equation}\label{fieldstrengths}
\begin{aligned}
K^I_{\mu\nu\rho\sigma}&:=-\frac{1}{2}\left(\nabla_\rho \nabla_{[\mu} h_{\nu] \sigma}^I-\nabla_\sigma \nabla_{[\mu} h_{\nu] \rho}^I+\nabla_\mu \nabla_{[\rho} h_{\sigma] \nu}^I-\nabla_\nu \nabla_{[\rho} h_{\sigma] \mu}^I\right) \\
&\quad +\frac{\sigma}{L^2}\left(g_{\rho[\mu} h_{\nu] \sigma}^I-g_{\sigma[\mu} h_{\nu] \rho}^I\right) \,,\\
\F_{\lambda \mu \nu}^{\Delta}&:=2\, \nabla_{[\lambda} k_{\mu] \nu}^{\Delta} \,,\qquad\qquad F^a_{\mu\nu}:=2\,\nabla_{[\mu}A^a_{\nu]} \,.
\end{aligned}
\end{equation}

The Noether procedure, or its generalization given in \cite{Berends:1984rq}, 
consists in the construction of interactions, perturbatively 
in a set of deformation parameters, under the requirement of maintaining the 
number of gauge symmetries.
No other restrictions are made a priori, although eventually we will set limits 
on the number of derivatives that may appear in the gauge transformations and in the 
Lagrangian, therefore ensuring locality. 
We also assume covariance of the deformation 
under the (A)dS background isometry algebra, as explained in \cite{Boulanger:2018fei}
to which we refer fore more details. 
As stated earlier, we actually consider the reformulation (and further 
generalization) of the deformation procedure of \cite{Berends:1984rq} 
using the BRST-BV cohomological approach spelled out in 
\cite{Barnich:1993vg,Henneaux:1997bm}.

We begin by defining an enlarged field content through the introduction of ghosts 
$\xi_\mu^I$, $\chi^{\Delta}$ and $C^a$ respectively associated with the gauge 
parameters $\epsilon_\mu^I, \epsilon^{\Delta}$ and $\epsilon^a$. 
We also introduce antifields and antighosts, collectively denoted by 
$\{\Phi_\Xi^*\}:=\{h_I^{* \mu \nu}, k_{\Delta}^{* \mu \nu}, A_a^{* \mu}, 
\xi_I^{* \mu}, \chi_{\Delta}^*, C_a^*\}$. 
These are canonically conjugate to the fields and ghosts, collectively 
denoted $\{\Phi^\Xi\}:=\{h_{\mu \nu}^I, k_{\mu \nu}^{\Delta}, A_\mu^a,$ $\xi_\mu^I, \chi^{\Delta}, C^a\}$, through the BV antibracket 
\begin{equation}
(A, B):=\frac{\delta^R A}{\delta \Phi^\Xi} \frac{\delta^L B}{\delta \Phi_\Xi^*}-\frac{\delta^R A}{\delta \Phi_\Xi^*} \frac{\delta^L B}{\delta \Phi^\Xi} \;,
\end{equation}
for any local functionals $A$ and $B$, and where we use De Witt's condensed 
notations for summations over repeated indices that imply integration over 
spacetime.
Note that ghosts and antifields are Grassmann-odd variables in the present 
context with only bosonic fields and gauge symmetries.

The fundamental object of interest in this formalism is the BV functional $W[\Phi^\Xi,\Phi^*_\Xi]$, which encodes all information about the interaction vertices and gauge structure of the theory. At the free field level it reads 
\begin{equation}
W_0=S_0+\int \rmd^D x \sqrt{-g}\left[h_I^{* \mu \nu}\left(2\, \nabla_{(\mu} \xi^I_{\nu)}\right)+k_{\Delta}^{* \mu \nu}\left(\nabla_\mu \nabla_\nu \chi^{\Delta}-\frac{\sigma}{L^2} \bar{g}_{\mu \nu} \chi^{\Delta}\right)+A_a^{* \mu} \nabla_\mu C^a\right] \,.
\end{equation}
In our conventions, the antifields $h_I^{* \mu \nu}\,$, 
$k_{\Delta}^{* \mu \nu}$ and $A_a^{* \mu}$ are tensors and not tensorial 
densities.

The consistency of the theory hinges on the invariance of the action under gauge symmetries and the existence of a consistent algebraic structure for the latter. In the BRST-BV formalism this is compactly enforced by the classical master equation
\begin{equation}
\left(W, W\right)=0 \,,
\end{equation}
and it is easy to verify, through the use of the free-theory Noether identities, that $W_0$ indeed satisfies this equation.

The reformulation of the Noether procedure continues with the perturbative expansion of the BV functional,
\begin{equation}
W=W_0+W_1+W_2+\ldots \,,
\end{equation}
where, in our set-up, $W_1$ is cubic in the fields and antifields, $W_2$ is quartic, and so on. The master equation is to be solved in perturbation theory so as to determine the most general $W_1$, $W_2$ and so on. Thus one determines first $W_1$ by solving $(W_0,W_1)=0$, next $W_2$ by solving $(W_0,W_2)=-\frac{1}{2}(W_1,W_1)$, and so on.

It has proved extremely useful to recast the procedure in the form of a 
cohomological problem \cite{Barnich:1993vg,Barnich:1994db,Barnich:1994mt,Henneaux:1997bm}. 
To this end one defines the BRST differential $s$, here given by 
$s\, \bullet:=\left(W_0, \bullet\right)$ as we are interested in deformations of 
a free theory.\footnote{In particular, note that in our conventions we have 
$s\,\Phi^\Xi(x) = -\frac{1}{\sqrt{-g}}\,\frac{\delta^R W_0}{\delta \Phi^*_\Xi(x)}\,$
and 
$s\,\Phi^*_\Xi(x) = \frac{1}{\sqrt{-g}}\,\frac{\delta^R W_0}{\delta \Phi^\Xi(x)}\,$.}
The first-order deformation $W_1$ should therefore satisfy 
$sW_1 = 0\,$.
Since $s$ is nilpotent, $s^2=0$, it follows that any $s$-exact 
contribution $sB$ to $W_1$ (where $B$ is a local functional) is a trivial 
solution of the master equation to that order.
In fact, such a solution must be discarded since it can be shown to 
correspond to a deformation generated from the free theory by trivial 
redefinition of the fields and gauge parameters. 
To sum up, it can be shown that a non-trivial deformation $W_1$ should 
satisfy $sW_1=0$ and should not be of the form $W_1=sB$ for a local 
function $B\,$, so that non-trivial cubic deformations are characterized 
by the cohomology of $s$ in the space of local functionals with ghost 
number zero. The ghost number (denoted by ${\rm gh}$) 
is a useful grading for the purpose of organizing the classification of 
solutions to the master equation. In the same vein it is helpful to 
also define the gradings called \textit{pure ghost number} (${\rm puregh}$ for short), 
and \textit{antifield number} (${\rm antifld}$ for short). 
The rationale for introducing these numbers will become clearer in the following, 
and we refer the reader to \cite{Barnich:2000zw,Boulanger:2000rq} for more complete 
explanations. The values of these gradings for the variables considered in this paper 
are given below in Table \ref{tab:gradings}.

Another useful ingredient is the decomposition of the BRST differential into 
$s=\gamma+\delta$. Here $\gamma$ is a differential which acts on the fields 
in the form of a gauge transformation in terms of the ghosts; 
the differential $\delta$ acts on the antifields to produce the linear equations 
of motion and corresponding Noether identities in terms of the fields and 
antifields, respectively. 
The actions of $\gamma$ and $\delta$ are
explicitly shown in Table \ref{tab:gradings} for the system under consideration. 
Once the actions of $\gamma$ and $\delta$ are known through 
Table \ref{tab:gradings} on the fields $\Phi^\Xi$ and antifields $\Phi^*_\Xi\,$, 
we extend their actions on the jet space of the fields, antifields and 
all their derivatives by asking $\gamma$ and $\delta$ to be derivations 
that anticommute with the total exterior differential $d\,$. In our conventions
for the antifields, note that one has 
$(A^a_\mu(x),A^{*\nu}_b(y))=\frac{1}{\sqrt{-g}}\,\delta^a_b\delta^\mu_\nu\delta^D(x-y)$ 
where $\delta^D(x-y)$ is the Dirac 
delta density obeying $\int d^Dx \,\delta^D(x-y)f(y)= f(x)\,$. 
We also note that, on top of the relations $\gamma^2=0=\delta^2\,$, one has 
$\gamma\delta=-\delta\gamma$. In our context the cohomology of $\gamma$ 
is easy to work out and will be extensively used in our analysis:
\begin{equation} \label{eq:gamma cohomology}
H(\gamma) \cong\left\{f\left([K_{\mu \nu\rho \sigma}^I],\left[\F_{\lambda \mu \nu}^{\Delta}\right],\left[F_{\mu \nu}^a\right], \xi_\mu^I, \nabla_{[\mu} \xi_{\nu]}^I, \chi^{\Delta}, \nabla_\mu \chi^{\Delta}, C^a,\left[\Phi_\Xi^*\right]\right)\right\} \,,
\end{equation}
where $f$ is an arbitrary function of the arguments shown, and by square 
brackets we mean the variable and all its (A)dS covariant derivatives.

\begin{table}[ht]
\centering
\begin{tabular}{|c||c|c|c|c||c|c|}
\hline & $|\bullet|$ & gh & puregh & antifld & $\gamma\,\bullet$ & $\delta\,\bullet$ \\
\hline \hline $h_{\mu \nu}^I$ & 0 & 0 & 0 & 0 & $2 \nabla_{(\mu} \xi^I_{\nu)}$ & 0 \\
\hline $k_{\mu \nu}^{\Delta}$ & 0 & 0 & 0 & 0 & $\nabla_\mu \nabla_\nu \chi^{\Delta}-\frac{\sigma}{L^2} g_{\mu \nu} \chi^{\Delta}$ & 0 \\
\hline $A_\mu^a$ & 0 & 0 & 0 & 0 & $\nabla_\mu C^a$ & 0 \\
\hline $\xi_\mu^I$ & 1 & 1 & 1 & 0 & 0 & 0 \\
\hline $\chi^{\Delta}$ & 1 & 1 & 1 & 0 & 0 & 0 \\
\hline $C^a$ & 1 & 1 & 1 & 0 & 0 & 0 \\
\hline $h_I^{* \mu \nu}$ & 1 & $-1$ & 0 & 1 & 0 & $\mathcal{E}^{\mu\nu}_I$ \\
\hline $k_{\Delta}^{* \mu \nu}$ & 1 & $-1$ & 0 & 1 & 0 & $\mathcal{E}^{\mu\nu}_\Delta$ \\
\hline $A_a^{* \mu}$ & 1 & $-1$ & 0 & 1 & 0 & $\mathfrak{g}_{a b}\,\nabla_\nu F^{b \nu \mu}$ \\
\hline $\xi_I^{* \mu}$ & 0 & $-2$ & 0 & 2 & 0 & $-2 \nabla_\nu h_I^{* \mu \nu}$ \\
\hline $\chi_{\Delta}^*$ & 0 & $-2$ & 0 & 2 & 0 & $\nabla_\mu \nabla_\nu k_{\Delta}^{* \mu \nu}-\frac{\sigma}{L^2} k_{\Delta}^*$ \\
\hline $C_a^*$ & 0 & $-2$ & 0 & 2 & 0 & $-\nabla_\mu A_a^{* \mu}$ \\
\hline
\end{tabular}
\caption{Gradings of fields and antifields, and the action of $\gamma$ and $\delta$ for the system studied in this paper. Here $|\bullet|$ denotes the Grassmann parity. For brevity we omit the explicit expressions of the equations of motion for the spin-2 fields, $\mathcal{E}^{\mu\nu}_I$ and $\mathcal{E}^{\mu\nu}_\Delta$, which may be straightforwardly inferred from $S_0$ (see also Appendix \ref{sec:appendix}).}
\label{tab:gradings}
\end{table}

%%%%%%%%%%%%%%%%%%%%%%%%%%%%
%%%%%%%%%%%%%%%%%%%%%%%%%%%%

\section{First order deformations: cubic vertices} \label{sec:firstorder}

In this section we consider the classification of non-trivial solutions to the master equation at first order in the deformation procedure,
\begin{equation} \label{eq:master eq W1}
s W_1=0 \,.
\end{equation}
Following Refs.\ \cite{Barnich:1994db,Barnich:1994mt,Boulanger:2000rq,Boulanger:2001he}, it proves useful to expand the BV functional $W_1$ in terms of local functionals with definite antifield number,
\begin{equation}
W_1=\int \rmd^D x \sqrt{-g}\left(a_0+a_1+a_2\right) \,,
\end{equation}
where ${\rm antifld}(a_n)=n$. As we are focusing on cubic deformations, it is easy to see that one cannot write terms with antifield number higher than 2 and ghost number zero. Each $a_n$ plays a specific role in the gauge structure of the theory: $a_2$ encodes the deformations of the gauge algebra (in particular, $a_2=0$ means that the algebra is Abelian); $a_1$ characterizes the deformations of the gauge transformations; and $a_0$ is nothing but the set of cubic vertices that we seek.

From the master equation \eqref{eq:master eq W1} we infer the following descent equations:
\begin{align}
\gamma a_2 & =0 \,,\label{eq:descent1a} \\
\delta a_2+\gamma a_1 & =\nabla_\mu j_1^\mu \,, \label{eq:descent1b} \\
\delta a_1+\gamma a_0 & =\nabla_\mu j_0^\mu \,. \label{eq:descent1c}
\end{align}
Here $j_n^{\mu}$ is some local vector 
with ${\rm antifld}(j_n^{\mu})=n$. 
Note that our perturbative assumption implies that a $j_2^{\mu}$ term does 
not exist with the required quantum numbers (antifield number 2 and ghost 
number 1). In fact, it is a general result, see \cite{Barnich:1994mt,Boulanger:2000rq}, 
that from the equation 
$\gamma a_k + \nabla_\mu j^{\mu}=0\,$ with $k>0$, one can redefine 
away $j^\mu$.

\subsection{Gauge algebra}

We start by determining the general solution of Eq.\ \eqref{eq:descent1a}. 
Non-trivial solutions correspond to the elements of the cohomology of 
$\gamma$ at antifield number 2 and ghost number zero, and must be scalars 
under the (A)dS isometries and parity-even. A complete basis of solutions 
is given by 
\begin{equation} \label{eq:a2 list}
\begin{aligned}
a_2^{\rm (EH)} & =\xi_I^{* \mu} \xi^{J \nu} \nabla_{[\mu} \xi_{\nu]}^k g^I{ }_{JK}, 
& a_2^{\rm (YM)} & =\frac{1}{2} C_a^* C^b C^c f^a{ }_{b c}, \\
a_2^{\rm (P M 1)} 
& =\chi_{\Delta}^* \chi^{\Omega} \chi^{\Gamma} \,m^{\Delta}{ }_{\Omega \Gamma}, 
& a_2^{\rm (P M 2)} & =\chi_{\Delta}^* \nabla_\mu \chi^{\Omega} \nabla^\mu \chi^{\Gamma} 
\,n^{\Delta}{ }_{\Omega \Gamma}, \\
a_2^{(1)} & =\xi_I^{* \mu} \xi_\mu^J \chi^{\Delta} \,f_{(1)}{ }^I{ }_{J \Delta}, 
& a_2^{(2)} & =\xi_I^{* \mu} \xi_\mu^J C^a \,f_{(2)}{ }^I{ }_{J a}, \\
a_2^{(3)} & =\xi_I^{* \mu} \nabla_{[\mu} \xi_{\nu]}^J \nabla^\nu \chi^{\Delta} 
\,f_{(3)}{ }^I{ }_{J \Delta}, & 
a_2^{(4)} & =\xi_I^{* \mu} C^a \nabla_\mu \chi^{\Delta} \,f_{(4)}{ }^I{ }_{a \Delta}, 
\\
a_2^{(5)} & =\xi_I^{* \mu} \chi^{\Delta} \nabla_\mu \chi^{\Omega} \,
f_{(5)}{ }^I{ }_{\Delta \Omega}, 
& a_2^{(6)} & =\chi_{\Delta}^* \chi^{\Omega} C^a \,f_{(6)}{ }^{\Delta}{ }_{\Omega a}, 
\\
a_2^{(7)} & =\chi_{\Delta}^* \nabla^\mu \chi^{\Omega} \xi_\mu^I \,
f_{(7)}{ }^{\Delta}{ }_{\Omega I}, & 
a_2^{(8)} & =\chi_{\Delta}^* C^a C^b \,f_{(8)}{ }^{\Delta}{ }_{a b}, \\
a_2^{(9)} & =\chi_{\Delta}^* \xi_\mu^I \xi^{J \mu} \,f_{(9)}{ }^{\Delta}{ }_{I J}, & 
a_2^{(10)} & =\chi_{\Delta}^* \nabla_{[\mu} \xi_{\nu]}^I 
\nabla^{[\mu} \xi^{\nu] J}{ }\,f_{(10)}{ }^{\Delta}{ }_{I J}, \\
a_2^{(11)} & =C_a^* C^b \chi^{\Delta}\, f_{(11)}{ }^a{ }_{b \Delta}, & a_2^{(12)} & =C_a^* \chi^{\Delta} \chi^{\Omega} \,f_{(12)}{ }^a{ }_{\Delta \Omega}, \\
a_2^{(13)} & =C_a^* \nabla_\mu \chi^{\Delta} \nabla^\mu \chi^{\Omega}{ }\,
f_{(13)}{ }^a{ }_{\Delta \Omega}, & 
a_2^{(14)} & =C_a^* \nabla^\mu \chi^{\Delta} \xi_\mu^I \,f_{(14)}{}^a{}_{\Delta I}, \\
a_2^{(15)} & =C_a^* \xi_\mu^I \xi^{J \mu}\,f_{(15)}{ }^a{ }_{I J}, 
& a_2^{(16)} & =C_a^* \nabla_{[\mu} \xi_{\nu]}^I \nabla^{[\mu} \xi^{\nu] J} 
\,f_{(16)}{ }^a{ }_{I J} \,.
\end{aligned}
\end{equation}
The structure constants appearing in these expressions satisfy some symmetrization constraints due to the Grassmann parity of the ghosts,
\begin{equation}
\begin{aligned}
& f^a{ }_{b c}=f^a{ }_{[b c]}, \quad m^{\Delta}{ }_{\Omega \Gamma}=m^{\Delta}{ }_{[\Omega \Gamma]}, \quad n^{\Delta}{ }_{\Omega \Gamma}=n^{\Delta}{ }_{[\Omega \Gamma]}, \quad f_{(8)}{ }_{a b}=f_{(8)}{ }_{[a b]} \,, \\
& f_{(9)}{ }^{\Delta}{ }_{IJ}=f_{(9)}{ }^{\Delta}{ }_{[IJ]}, \quad f_{(10)}{ }^{\Delta}{ }_{IJ}=f_{(10)}{ }^{\Delta}{ }_{[IJ]}, \quad f_{(12)}{ }^a_{\Delta \Omega}=f_{(12)}{ }^a{ }_{[\Delta \Omega]} \,, \\
& f_{(13)}{ }^a{}_{\Delta \Omega}=f_{(13)}{ }^a{}_{[\Delta \Omega]}, \quad f_{(15)}{ }^a{ }_{IJ}=f_{(15)}{ }^a{ }_{[IJ]}, \quad f_{(16)}{ }^a{ }_{IJ}=f_{(16)}{ }^a{ }_{[IJ]} \,. \\
&
\end{aligned}
\end{equation}
In the list \eqref{eq:a2 list}, $a_2^{\rm (EH)}$ is the massless multi-graviton, 
called Einstein-Hilbert deformation \cite{Boulanger:2000rq}, 
$a_2^{\rm (YM)}$ is the usual Yang-Mills deformation \cite{Barnich:1994mt}, 
and $a_2^{\rm (PM1)}$, $a_2^{\rm (PM2)}$ are the unique candidate deformations 
of the PM spin-2 gauge algebra without additional fields \cite{Boulanger:2019zic}. 
The rest of the terms correspond to mixings among the gauge variations for different 
field types. We emphasize that this list is complete insofar as cubic deformations 
are concerned, without any restriction on the number of derivatives.

Not all among the candidates in \eqref{eq:a2 list} are admissible at first 
order in the deformation procedure. Indeed the second descent equation, 
Eq.\ \eqref{eq:descent1b}, states that $\delta a_2$ must be $\gamma$-exact modulo 
total derivatives. We find the following linear combination of candidates to be 
unobstructed at this order:
\begin{equation} \label{eq:a2 lift}
\begin{aligned}
a_2&=\kappa \,a_2^{\rm (E H)} +g_{\rm (Y M)} \,a_2^{\rm (Y M)}
+\kappa_{(2)}\, a_2^{(2)}+\kappa_{(5)} \,a_2^{(5)}+\kappa_{(6)} \,a_2^{(6)}
+\kappa_{(7)}\, a_2^{(7)} \\
&\quad +\kappa_{(13)}\,\left(a_2^{(13)}-\frac{\sigma}{L^2} 
a_2^{(12)}\right)+\kappa_{(16)}\,\left(a_2^{(16)}-\frac{\sigma}{L^2} a_2^{(15)}\right) \,,
\end{aligned}
\end{equation}
where $\kappa$, $g_{\rm YM}$ and $\kappa_{(i)}$ are arbitrary coupling coefficients,\footnote{These constants are of course just for bookkeeping since one may choose to absorb them into the structure constants.} in addition to the following linear constraints on the structure constants:
\begin{equation}
f_{(12)}{ }^a{}_{\Delta \Omega}=f_{(13)}{ }^a{}_{\Delta \Omega} \,, \quad f_{(15)}{ }^a{ }_{IJ}=f_{(16)}{ }^a{ }_{IJ} \,, \quad f_{(5)}{ }^I{}_{\Delta \Omega}=f_{(5)}{ }^I{}_{(\Delta \Omega)} \,.
\end{equation}

\subsection{Gauge symmetries}

We have already identified in Eq.\ \eqref{eq:a2 lift} the deformation $a_2$ which is not unobstructed at this order in the procedure. A particular solution of the descent equation for $a_1$ is given by
\begin{equation}
\begin{aligned}
\hat{a}_1&=\kappa \,\hat{a}_1^{\rm (E H)} +g_{\rm (Y M)}\, 
\hat{a}_1^{\rm (Y M)}+\kappa_{(2)} \,\hat{a}_1^{(2)}
+\kappa_{(5)}\, \hat{a}_1^{(5)}+\kappa_{(6)}\, \hat{a}_1^{(6)}
+\kappa_{(7)}\, \hat{a}_1^{(7)} \\
&\quad +\kappa_{(13)}\,\hat{a}_1^{(12-13)}
+\kappa_{(16)}\,\hat{a}_1^{(15-16)} \,,
\end{aligned}
\end{equation}
where each $\hat{a}_1^{(i)}$ solves $\delta a_2^{(i)}+\gamma \hat{a}_1^{(i)}=(\mbox{total derivative})$, with the the $a_2^{(i)}$ as given in \eqref{eq:a2 lift}. The explicit expressions are given by
\begin{equation} \label{eq:a1 list}
\begin{aligned}
\hat{a}_1^{\rm(EH)} & =h_I^{*\mu \nu}\left(h_{\mu \sigma}^J \nabla_\nu \xi^{K \sigma}-\nabla_\mu h_{\nu \sigma}^K \xi^{J \sigma}+\nabla_\sigma h_{\mu \nu}^K \xi^{J \sigma}\right) g^I{ }_{J K} \,, \\
\hat{a}_1^{\rm(YM)} & =A_a^{*\mu} A_\mu^b C^c f^a{ }_{b c} \,, \\
\hat{a}_1^{(2)} & =h_I^{*\mu \nu}\left(\frac{1}{2} h_{\mu \nu}^J C^a-A_\mu^a \xi_\nu^J\right) f_{(2)}{ }^I{ }_{J a} \,, \\
\hat{a}_1^{(5)}&= h_I^{*\mu \nu} k_{\mu \nu}^{\Delta} \chi^{\Omega} f_{(5)}{ }^I{ }_{\Delta \Omega} \,, \\
\hat{a}_1^{(6)} & =k_{\Delta}^{*\mu \nu}\left(k_{\mu \nu}^{\Omega} C^a-\nabla_\mu A_\nu^a \chi^{\Omega}-2 A_\mu^a \nabla_\nu \chi^{\Omega}\right) f_{(6)}{ }^{\Delta}{ }_{\Omega a} \,, \\
\hat{a}_1^{(7)}&=  k_{\Delta}^{*\mu \nu}\bigg(\nabla_\mu k^{\Omega}{ }_{\nu \sigma} \xi^{I \sigma}+2 k_{\mu \sigma}^{\Omega} \nabla_\nu \xi^{I \sigma}-\nabla_\mu h_{\nu \sigma}^I \nabla^\sigma \chi^{\Omega} \\
& \quad+\frac{1}{2} \nabla_\sigma h_{\mu \nu}^I \nabla^\sigma \chi^{\Omega}-\frac{\sigma}{L^2} h_{\mu \nu}^I \chi^{\Omega}\bigg) f_{(7)}{ }^{\Delta}{ }_{\Omega I} \,, \\
\hat{a}_1^{(12-13)}& =2 A_a^{*\mu} k_{\mu \nu}^{\Delta} \nabla^\nu \chi^{\Omega} f_{(13)}{ }^a{ }_{\Delta \Omega} \,, \\
\hat{a}_1^{(15-16)}&= A_a^{*\mu} \bigg(2\nabla^{\nu} h^{\sigma}{ }_\mu{ }^I \nabla_{[\nu} \xi_{\sigma]}^J -\frac{\sigma}{L^2}  h_{\mu \nu}^I \xi^{J \nu} \bigg)f_{(16)}{ }^a{ }_{IJ} \,.
\end{aligned}
\end{equation}

Notice that the PM deformations, $a_2^{\rm (PM1)}$ and $a_2^{\rm (PM2)}$, 
are obstructed at this order \cite{Boulanger:2019zic}. 
To the above particular solution one must add the general solution 
$\bar{a}_1$ of the homogeneous equation
\beq
\gamma\bar{a}_1=\nabla_{\mu}\bar{\jmath}_1^{\,\mu} \,.
\eeq
Since antigh$(\bar{a}_1)=1>0$, general results 
\cite{Barnich:1994mt,Boulanger:2000rq}
show that one can absorb away 
the right-hand side, yielding $\gamma\bar{a}_1=0\,$.
These correspond to Abelian, but non-linear, deformations of the gauge symmetries, 
sometimes referred to as Chapline-Manton type deformations in analogy to the theories 
studied in \cite{Chapline:1982ww,Henneaux:1997ha,Freedman:1980us}. At cubic order, 
$\bar{a}_1$ contains one power of the curvatures or any number of derivatives thereof 
(cf.\ Eq.\ \eqref{eq:gamma cohomology}). In order to have a bounded number of solutions, 
it is therefore necessary at this stage to set a limit on the number of derivatives. 
Keeping in mind that our aim is to obtain general cubic vertices with at most two 
derivatives, we are thus led to consider at most four derivatives in $a_1$ (because 
$\gamma$ increases the number of derivatives by at most two, cf.\ Table 
\ref{tab:gradings}). With this restriction, the classification of Abelian solutions 
$\bar{a}_1$ is straightforward, although quite cumbersome. 
The full list may be found in~\cite{lucas-thesis}.

As before, our main objective is to determine which among all these candidate $a_1$ 
are unobstructed in the next descent equation, Eq.\ \eqref{eq:descent1c}. 
At this stage we note that our assumption on the number of derivatives allowed in the 
space of solutions presents an issue: it may occur that a candidate $a_1$ is 
unobstructed, but the corresponding vertex $a_0$ contains strictly more than two 
derivatives. However, one must check that the higher-derivative terms are not spurious, 
in the sense that the solution is actually equivalent to a two-derivative vertex via a 
field redefinition. In order not to miss such solutions in our classification, it proves 
useful to include in our list of $a_1$ trivial terms which are $\gamma$-exact, since then 
the corresponding $a_0$ will be $\delta$-exact (i.e.\ it can be removed by a field 
redefinition). The list of $\gamma$-exact $a_1$ candidates with the required number of 
derivatives is also given in \cite{lucas-thesis}.

\subsection{Cubic vertices}

In this subsection we present the list of solutions for the cubic vertices consistent with the last descent equation. At this stage further constraints appear on the structure constants of the gauge algebra (and also, in some instances, on the spacetime dimension $D$), and we also indicate in each case the corresponding deformations of the gauge symmetries and algebra consistent with the existence of a two-derivative vertex.

\paragraph{Einstein-Hilbert coupling.} The deformation $a_1=\hat{a}_1^{\rm(EH)}$ is a consistent solution of Eq.\ \eqref{eq:descent1c} provided the structure constants satisfy
\begin{equation}
g_{IJK}=g_{(IJK)} \qquad\qquad \left(g_{IJK}:=g^L{ }_{JK}\mathfrak{g}_{IL}\right) \,.
\end{equation}
The corresponding vertex $a_0$ is the multi-graviton Einstein-Hilbert 
solution \cite{Boulanger:2000rq},
\begin{equation}
\begin{aligned}
a_0^{\rm(EH)}&= g_{IJK}\bigg[\frac{1}{2} h^{I \mu \nu} \nabla_\mu h_{\rho \sigma}^J \nabla_\nu h^{K \rho \sigma}-\frac{1}{2} h^{I \mu \nu} \nabla_\mu h^J \nabla_\nu h^K+h^{I \mu \nu} \nabla_\mu h^J \nabla^\sigma h_{\sigma \nu}^K \\
&\quad +h^{I \mu \nu} \nabla_\mu h_{\nu \sigma}^J \nabla^\sigma h^K-h^{I \mu \nu} \nabla_\sigma h_{\mu \nu}^J \nabla^\sigma h^K+\frac{1}{4} h^I \nabla_\mu h^J \nabla^\mu h^K+h^{I \mu \nu} \nabla_\sigma h_{\mu \nu}^J \nabla_\rho h^{K \rho \sigma} \\
&\quad -\frac{1}{2} h^I \nabla_\mu h^J \nabla_\nu h^{K \mu \nu}-2 h^{I \mu}{ }_\nu \nabla_\mu h_{\rho \sigma}^J \nabla^\rho h^{K \sigma \nu}-h^{I \mu}{ }_\nu \nabla_\sigma h_{\rho \mu}^J \nabla^\rho h^{K \sigma \nu} \\
&\quad +h^{I \mu}{ }_\nu \nabla_\sigma h_{\rho \mu}^J \nabla^\sigma h^{K \rho \nu}+\frac{1}{2} h^I \nabla_\sigma h_{\mu \nu}^J \nabla^\mu h^{K \nu \sigma}-\frac{1}{4} h^I \nabla_\sigma h_{\mu \nu}^J \nabla^\sigma h^{K \mu \nu} \\
&\quad +\frac{\sigma}{L^2}\bigg(\frac{1}{2} h^I h^J h^K+4 h^I{ }_\mu{ }^\nu h^J{ }_\nu{ }^\rho h^K{ }_\rho{ }^\mu-3 h^I h_{\mu \nu}^J h^{K \mu \nu}\bigg)\bigg] \,.
\end{aligned}
\end{equation}

\paragraph{Yang-Mills coupling.} Similarly the deformation 
$a_1=\hat{a}_1^{\rm(YM)}$ is a consistent solution of the master equation provided
\begin{equation}
f_{abc}=f_{[abc]} \qquad\qquad \left(f_{abc}:=\mathfrak{g}_{ad}\,f^d{ }_{bc}\right) \,.
\end{equation}
The vertex is given by
\begin{equation}
a_0^{\rm (YM)}=-\frac{1}{2}\, f_{abc}\, A_\mu^a A_\nu^b F^{c\mu \nu} \,.
\end{equation}
In the BRST-BV formalism the Yang-Mills system was first studied in 
\cite{Barnich:1994mt}. See also \cite{Boulanger:2018dau} 
for the case of massive Yang-Mills theory.

\paragraph{Partially massless spin-2 self-coupling.} We have seen already that the non-Abelian deformations of the PM spin-2 gauge algebra are obstructed. However, the Abelian term
\begin{equation}
\bar{a}_1^{\rm(PM)}:=c_{\Sigma \Delta \Omega} \,k^{* \Delta \mu \nu}\, 
\F_{\sigma \mu \nu}^{\Omega}\, \nabla^\sigma \chi^{\Sigma} \quad\in\; H(\gamma)
\end{equation}
leads to a consistent vertex provided the constants $c_{\Sigma \Delta \Omega}$ satisfy
\begin{equation}
c_{\Sigma \Delta \Omega}=c_{\Sigma(\Delta \Omega)} \,.
\end{equation}
Moreover, the spacetime dimension must be $D=4$ for the solution to exist. The vertex is given by \cite{Boulanger:2019zic}
\begin{equation}
a_0^{\rm(PM)}=\tfrac{1}{2}\,k_{\mu \nu}^{\Delta} J_{\Delta}^{\mu \nu}
\end{equation}
in terms of the gauge-invariant current
\begin{equation} \label{eq:current PM vertex}
\begin{aligned}
J_\Delta^{\mu \nu}&:= 2\,c_{\Sigma \Delta \Omega}\bigg(\F^{\Omega(\mu \mid \rho \sigma} \F^{\Sigma \mid \nu)}{ }_{\rho \sigma}-\F^{\Omega(\mu \mid} \F^{\Sigma \mid \nu)}+\F^{\Omega(\mu|\sigma| \nu)} \F_\sigma^{\Sigma} \\
&\quad -\frac{1}{4} g^{\mu \nu} \F^{\Omega \rho \sigma \lambda} \F_{\rho \sigma \lambda}^{\Sigma}+\frac{1}{2} g^{\mu \nu} \F_\lambda^{\Omega} \F^{\Sigma \lambda}\bigg) \,.
\end{aligned}
\end{equation}
A priori, this current should only satisfy the partially-massless conservation law 
for consistency of the vertex,
but here it happens to obey the stronger constraints 
$\nabla_\mu J_\Delta^{\mu \nu}\approx 0\,$ (where the weak equality symbol $\approx$ stands for an equality that holds on the solutions of the free equations of motion), as well as  
$g_{\mu\nu}\,J_\Delta^{\mu \nu}= 0\,$ since $D=4$; 
see  \cite{Boulanger:2019zic} for more details.

\paragraph{Spin-1 gravitational coupling.} Another solution is given by the minimal gravitational coupling of spin-1 fields. This arises from the Abelian deformation of the gauge symmetry
\beq \label{eq:a1 vector-g}
\bar{a}_1^{(v-g)}=d_{Iab} A^{* a\mu}F^{b}{}_{\mu\nu}\xi^{I\nu} \,,
\eeq
with the requirement
\beq
d_{Iab}=d_{I(ab)} \,.
\eeq
The vertex reads
\begin{equation}
a_0^{(v-g)}=-\frac{1}{2}d_{Iab} h_{\mu \nu}^I \left(F_{\mu \sigma}^a F^b{ }_\nu{}^\sigma-\frac{1}{4} g_{\mu \nu} F_{\rho \sigma}^a F^{b \rho \sigma}\right) \,,
\end{equation}
in which we indeed recognize the standard minimal coupling of gravity 
with the spin-1 energy-momentum tensor of Maxwell's theory.

It may appear strange that the gravitational coupling of a field arises here from 
an Abelian deformation of the gauge symmetry. In fact, we can obtain the more 
standard Lie derivative transformation, given in our language by\footnote{More 
explicitly, the Lie derivative of the vector field is
\begin{equation*}
d_{Iab}A^{\star a\mu}(-\pounds_{\xi^I}A^b_{\mu})=-d_{Iab}A^{*a\mu}(\xi^{I\nu}\nabla_{\nu}A^{b}_{\mu}+\nabla_{\mu}\xi^{I\nu}A^{b}_{\nu}) \,.
\end{equation*}
This differs from $a_1^{\rm(Lie)}$ in \eqref{eq:a1 vector-Lie} by a total derivative. We have chosen the form given in \eqref{eq:a1 vector-Lie} in order to make the equivalence with \eqref{eq:a1 vector-g} more manifest. Notice also, incidentally, that $\hat{a}_1^{\rm(EH)}$ (cf.\ Eq.\ \eqref{eq:a1 list}) also agrees with the Lie derivative 
of the massless spin-2 field 
$\delta_\xi h_{\mu\nu} = h_{\mu\sigma}\nabla_{\nu}\xi^\sigma
+h_{\nu\sigma}\nabla_{\mu}\xi^\sigma + \xi^\sigma\,\nabla_{\sigma}h_{\mu\nu}\,$, 
modulo the redefinition $\xi_\mu \mapsto \xi_\mu -\tfrac{1}{2}\,h_{\mu\nu}\xi^\nu$ 
of the gauge parameter in the Abelian transformation law 
$\gamma h_{\mu\nu}=2\nabla_{(\mu}\xi_{\nu)}$.}
\begin{equation} \label{eq:a1 vector-Lie}
a_1^{\rm(Lie)}=d_{Iab}\left(\nabla_{\mu}A^{* a\mu}A^{b}{}_{\nu}\xi^{I\nu} +A^{*a\mu}F^{b}{}_{\mu\nu}\xi^{I\nu}\right) \,,
\end{equation}
and we observe that \eqref{eq:a1 vector-g} and \eqref{eq:a1 vector-Lie} differ by a trivial 
$\delta$-exact term (with a correspondingly trivial $\gamma$-exact gauge algebra term 
$a_2^{\rm(Lie)}=d_{Iab}C^{*a}\nabla_{\nu}C^b \xi^{I\nu}$ signaling a trival redefinition 
of the gauge parameters of Maxwell's theory). 
Therefore $\bar{a}_1^{(v-g)}$ is indeed equivalent to the expected geometric-type deformation.

\paragraph{Mixed spin-1 and partially massless spin-2 coupling.} Spin-1 and 
PM spin-2 fields may also interact through a geometric-looking coupling 
involving the spin-1 energy-mo\-mentum tensor,
\begin{equation}
\label{Chaplin-Manton-like}
a_0^{(v-{\rm PM})}=-e_{\Delta ab} k^{\Delta \mu \nu} \left(F_{\mu \sigma}^a F^b{ }_\nu{}^\sigma-\frac{1}{4} g_{\mu \nu} F_{\rho \sigma}^a F^{b \rho \sigma}\right) \,,
\end{equation}
where the constants $e_{\Delta ab}$ must satisfy
\begin{equation}
 e_{\Delta ab}=e_{\Delta (ab)} \,,
\end{equation}
and the spacetime dimension must be $D = 4$ for the solution to exist.
The corresponding deformation of the gauge transformation is Abelian and reads
\begin{equation}
\bar{a}_1^{(v-{\rm PM})}=e_{\Delta a b} A^{* \mu a} F_{\mu \nu}^b \nabla^\nu \chi^{\Delta} \,.
\end{equation}

\paragraph{Partially massless spin-2 gravitational coupling.} 
Next we consider couplings between PM and massless spin-2 fields. We will confirm here the results of Ref.\ \cite{Joung:2019wwf}. 
We derive first the non-Abelian deformation corresponding to the minimal 
coupling of PM fields to gravity. 
It turns out that, in order to have a consistent vertex starting from 
$a_2^{(5)}$ and $a_2^{(7)}$, a precise linear combination of them should be taken
with $f_{(5)}{ }^I{}_{\Delta \Omega}=f_{(7) \Delta \Omega}{ }^I$, 
yielding
\begin{equation}
a_2^{({\rm PM}-g)}=2a_2^{(7)} -\frac{(D-3)\sigma}{2 L^2}a_2^{(5)} \,,\qquad
f_{(5)}{ }^I{}_{\Delta \Omega}=f_{(7) \Delta \Omega}{ }^I\;,
\end{equation}
in terms of the $a_2^{(i)}$ deformations given in Eq.\ \eqref{eq:a2 list}. 
The above linear combination is necessary but not sufficient to produce a consistent 
vertex. The corresponding particular solution $\hat{a}_1$ still needs to be completed 
by the addition of a suitable Abelian solution $\bar{a}_1$ solution of $\gamma \bar{a}_1=0$, 
yielding 
\beq
a_1^{({\rm PM}-g)}= 2  \hat{a}_1^{(7)}+\frac{(D-3)\sigma}{L^2} \hat{a}_1^{(5)}+a_{I\Delta \Omega}\left(h^{*I\mu \nu} \F_{\sigma \mu \nu}^{\Delta} \nabla^\sigma \chi^{\Omega} +2  k^{*\Delta \mu \nu} \F_{\sigma \mu \nu}^{\Omega} \xi^{I \sigma}\right) \,,
\eeq
where the $\hat{a}_1^{(i)}$ were given above in Eq.\ \eqref{eq:a1 list}, and where 
\begin{equation}
f_{(5)}{ }^I{}_{\Delta \Omega}=f_{(7) \Delta \Omega}{ }^I=a^I{ }_{\Delta \Omega}=a^I{ }_{(\Delta \Omega)} \,.
\end{equation}
More explicitly, we find
\begin{equation}
\begin{aligned}
a_1^{({\rm PM}-g)}&= 2\,a_{I\Delta \Omega}\, k^{*\Delta \mu \nu}\bigg(
\xi^{I \sigma}\,\nabla_\sigma k_{\mu\nu}^{\Omega} +2\, k_{\mu \sigma}^{\Omega} \nabla_\nu 
\xi^{I \sigma}-\nabla_\mu h_{\nu \sigma}^I \nabla^\sigma \chi^{\Omega}
+\frac{1}{2}\, \nabla^\sigma \chi^{\Omega}\,\nabla_\sigma h_{\mu \nu}^I  \\
&\quad -\frac{\sigma}{L^2} h_{\mu \nu}^I \chi^{\Omega}\bigg) 
+a_{I \Delta \Omega}\,h^{*I \mu \nu}\bigg(\F_{\sigma \mu \nu}^{\Delta} \nabla^\sigma 
\chi^{\Omega}+\frac{(D-3)\sigma}{L^2}k_{\mu \nu}^{\Delta} \chi^{\Omega}\bigg) \,.
\end{aligned}
\end{equation}
Note that the first two terms correspond to the Lie-derivative transformation of 
the PM spin-2 field $k^\Delta_{\mu \nu}\,$.
Lastly we provide the expression for the cubic vertex, which we write in the form
\begin{equation}
a_0^{({\rm PM}-g)}=h_{\mu \nu}^I T_I^{\mu \nu} \,,
\label{geom_vertex1}
\end{equation}
with
\begin{equation}
\begin{aligned}
\label{geom_vertex2}
\,&T_{I}{}^{\mu\nu}:=a_{I\Delta \Omega}\bigg[2k^{\Delta\nu\alpha}\nabla_{\alpha}\nabla_{\beta}k^{\Omega\mu\beta}+2k^{\Delta\mu\alpha}\nabla_{\alpha}\nabla_{\beta}k^{\Omega\nu\beta}+\nabla_{\alpha}k^{\Omega}\nabla^{\alpha}k^{\Delta\mu\nu}+k^{\Delta}\square k^{\Omega\mu\nu}\\
&\quad 
-2k^{\Delta\mu\nu}\nabla_{\alpha}\nabla_{\beta}k^{\Omega\alpha\beta}-2k^{\Delta\alpha\beta}\nabla_{\beta}\nabla_{\alpha}k^{\Omega\mu\nu}+2k^{\Delta\mu\nu}\square k^{\Omega}-2\nabla_{\alpha}k^{\Delta\mu\nu}\nabla_{\beta}k^{\Omega\alpha\beta}\\
&\quad  -\frac{3}{2}k^{\Delta\nu\alpha}\square k^{\Omega\mu}{}_{\alpha}-\frac{3}{2}k^{\Delta\mu\alpha}\square k^{\Omega\nu}{}_{\alpha}+2\nabla_{\alpha}k^{\Omega\nu\beta}\nabla_{\beta}k^{\Delta\mu\alpha}-\nabla_{\beta}k^{\Omega\nu\alpha}\nabla^{\beta}k^{\Delta\mu}{}_{\alpha}\\
&\quad  -\nabla_{\alpha}k^{\Omega}\nabla^{\mu}k^{\Delta\nu\alpha}+\frac{3}{2}\nabla_{\beta}k^{\Omega\alpha\beta}\nabla^{\mu}k^{\Delta\nu}{}_{\alpha}-\frac{1}{2}\nabla_{\beta}k^{\Delta\nu\alpha}\nabla^{\mu}k^{\Omega\beta}{}_{\alpha}-2k^{\Delta\nu\alpha}\nabla^{\mu}\nabla_{\alpha}k^{\Omega}\\
&\quad  +\frac{3}{2}k^{\Delta\nu\alpha}\nabla^{\mu}\nabla_{\beta}k^{\Omega\beta}{}_{\alpha}+\frac{3}{2}k^{\Delta\alpha\beta}\nabla^{\mu}\nabla_{\beta}k^{\Omega\nu}{}_{\alpha}-k^{\Delta}\nabla^{\mu}\nabla_{\beta}k^{\Omega\nu\beta}-\nabla_{\alpha}k^{\Omega}\nabla^{\nu}k^{\Delta\mu\alpha}\\
&\quad  +\frac{3}{2}\nabla_{\beta}k^{\Omega\alpha\beta}\nabla^{\nu}k^{\Delta\mu}{}_{\alpha}-\frac{1}{2}\nabla_{\beta}k^{\Delta\mu\alpha}\nabla^{\nu}k^{\Omega\beta}{}_{\alpha}-2k^{\Delta\mu\alpha}\nabla^{\nu}\nabla_{\alpha}k^{\Omega}+\frac{3}{2}k^{\Delta\mu\alpha}\nabla^{\nu}\nabla_{\beta}k^{\Omega\beta}{}_{\alpha}\\
&\quad  +\frac{3}{2}k^{\Delta\alpha\beta}\nabla^{\nu}\nabla_{\beta}k^{\Omega\mu}{}_{\alpha}-k^{\Delta}\nabla^{\nu}\nabla_{\beta}k^{\Omega\mu\beta}-k^{\Delta\alpha\beta}\nabla^{\mu}\nabla^{\nu}k^{\Omega}{}_{\alpha\beta}+k^{\Delta}\nabla^{\nu}\nabla^{\mu}k^{\Omega}\\
&\quad  +2g^{\mu\nu}k^{\Delta\alpha\beta}\nabla_{\beta}\nabla_{\alpha}k^{\Omega}-3g^{\mu\nu}k^{\Delta\alpha\beta}\nabla_{\beta}\nabla_{\gamma}k^{\Omega\gamma}{}_{\alpha}-\frac{1}{2}g^{\mu\nu}(\nabla_{\beta}k^{\Omega})(\nabla^{\beta}k^{\Delta})\\
&\quad  +2g^{\mu\nu}\nabla_{\beta}k^{\Delta}\nabla_{\gamma}k^{\Omega\beta\gamma}+g^{\mu\nu}k^{\Delta}\nabla_{\gamma}\nabla_{\beta}k^{\Omega\beta\gamma}+g^{\mu\nu}k^{\Delta\alpha\beta}\square k^{\Omega}{}_{\alpha\beta}-g^{\mu\nu}k^{\Delta}\square k^{\Omega}\\
&\quad  -\frac{1}{2}g^{\mu\nu}\nabla_{\beta}k^{\Omega\alpha\gamma}\nabla_{\gamma}k^{\Delta\beta}{}_{\alpha}+\frac{1}{2}g^{\mu\nu}\nabla_{\gamma}k^{\Omega\alpha\beta}\nabla^{\gamma}k^{\Delta}{}_{\alpha\beta}-\frac{3}{2}g^{\mu\nu}\nabla_{\alpha}k^{\Delta\alpha\beta}\nabla^{\gamma}k^{\Omega}{}_{\gamma\beta}\\
&\quad  +\frac{\sigma}{L^2}\bigg(4k^{\Delta\mu\nu}k^{\Omega}+(2D+3)k^{\Delta\mu\alpha}k^{\Omega\nu}{}_{\alpha}+\frac{D+7}{2}g^{\mu\nu}k^{\Delta\alpha\beta}k^{\Omega}{}_{\alpha\beta}+\frac{D-7}{2}g^{\mu\nu}k^{\Delta}k^{\Omega}\bigg)\bigg] \,.
\end{aligned}
\end{equation}
This current satisfies $\nabla_{\mu}T_I^{\mu \nu}=0$ on the solutions of the 
PM free-field equations of motion (see Appendix \ref{sec:appendix}), 
implying the consistency of the vertex under 
linearized diffeomorphisms. Less trivial is to check the consistency under PM gauge 
transformations since $T_I^{\mu \nu}$ is not itself gauge invariant.

\paragraph{Partially massless spin-2 non-geometric coupling.} Finally we note the 
existence of an exotic coupling between massless and PM spin-2 fields, first identified 
in Ref.\ \cite{Joung:2019wwf}, and dubbed \textit{non-geometric} due to the fact that it is 
Abelian ($a_2=0$). The deformation of the gauge symmetry is encoded in
\begin{equation}
\begin{aligned}
\bar{a}_1^{({\rm non-geo})}&= b_{I\Delta \Omega}\,k^{*\Delta \mu \nu}\bigg[ \F_{\sigma \mu \nu}^{\Omega} \xi^{I \sigma}-\frac{\sigma L^2}{2(D-2)} \nabla_\mu \left(\F_{\rho \sigma \nu}^{\Omega} \nabla^{[\rho} \xi^{\sigma] I}\right)\bigg] \,,
\end{aligned}
\end{equation}
with a restriction on the structure constants given by
\begin{equation}
b_{I\Delta \Omega}=b_{I(\Delta \Omega)} \,.
\end{equation}
This $a_1$ can be lifted through \eqref{eq:descent1c} to furnish the vertex
\begin{equation}
\label{non-geom_vertex1}
a_0^{({\rm non-geo})}=\frac{1}{2}\, h_{\mu \nu}^I J_I^{\mu \nu} \,,
\end{equation}
with the gauge-invariant current-like tensor
\begin{equation}
\begin{aligned}
\label{non-geom_vertex2}
J_I^{\mu \nu}&:= b_{I\Omega \Sigma}\bigg(\F^{\Omega(\mu \mid \rho \sigma} \F^{\Sigma \mid \nu)}{ }_{\rho \sigma}-\F^{\Omega(\mu \mid} \F^{\Sigma \mid \nu)}+\F^{\Omega(\mu|\sigma| \nu)} \F_\sigma^{\Sigma} \\
&\quad -\frac{1}{4} g^{\mu \nu} \F^{\Omega \rho \sigma \lambda} \F_{\rho \sigma \lambda}^{\Sigma}+\frac{1}{2} g^{\mu \nu} \F_\lambda^{\Omega} \F^{\Sigma \lambda}\bigg) \,.
\end{aligned}
\end{equation}
This current is actually identical (up to a coefficient that one may choose 
to factor out) to the one found in the PM self-coupling, 
Eq.\ \eqref{eq:current PM vertex}.
In that context, $J_\Delta^{\mu \nu}$ has a physical interpretation as it is 
related to a Noether current associated with a global symmetry of the free PM theory, 
a property which holds only in $D=4$ dimensions \cite{Boulanger:2019zic}. 
Similarly, in the present case, we have a Noether current $\mathcal{J}^\mu_{\Delta I}:=\sqrt{-g}J^{\mu\nu}_\Delta \overline{\epsilon}_{I\nu}\,$, where by definition $\overline{\epsilon}^I_\mu$ is a Killing vector of the background (A)dS space obeying $\nabla_{(\mu}\epsilon^I_{\nu)}=0\,$. 
Given that $\nabla_{\mu}J^{\mu\nu}_\Delta\approx0\,$, i.e., the (A)dS covariant divergence of $J^{\mu\nu}_\Delta$ vanishes on the solutions of the 
free equations of motion (cf.\ Appendix \ref{sec:appendix}), 
it follows immediately that $\partial_{\mu}\mathcal{J}^\mu_{\Delta I}\approx0\,$, without any restriction on the spacetime dimension. 
The corresponding global symmetry transformation law can be read off from the above expression for $\bar{a}_1^{({\rm non-geo})}$, where the gauge parameter $\epsilon^I_\mu$ 
must be replaced by 
the Killing parameter  $\overline{\epsilon}^I_\mu\,$.

%%%%%%%%%%%%%%%%%%%%%%%%%%%%
%%%%%%%%%%%%%%%%%%%%%%%%%%%%

\section{Second order deformations: quadratic constraints} \label{sec:secondorder}

In this Section we investigate the master equation at second order. We will not carry out a classification of the second-order BV functional $W_2$, but rather use the master equation as a consistency condition on the first order deformations identified in the previous Section. This will lead to a set of quadratic constraints on the structure constants, i.e.\ generalized Jacobi identities.

As before we split the second-order BV functional by antifield number,
\beq
W_2=\int \rmd^D x \sqrt{-g}\left(b_0+b_1+b_2\right) \,.
\eeq
It is again easy to verify that the expansion stops at antifield number 2. The master equation at this order in the perturbative analysis reads
\begin{equation}
s W_2=-\frac{1}{2}\left(W_1, W_1\right) \,,
\end{equation}
or, in terms of the local functions $a_n$ and $b_n$,
\begin{align}
\gamma b_2 & =-\frac{1}{2}\left(a_2, a_2\right)+\nabla_\mu t_2^\mu \,, \label{eq:descent2a}\\ 
\delta b_2+\gamma b_1 & =-\frac{1}{2}\left(a_1, a_1\right)-\left(a_2, a_1\right)+\nabla_\mu t_1^\mu \,, \label{eq:descent2b} \\
\delta b_1+\gamma b_0 & =-\left(a_1, a_0\right)+\nabla_\mu t_0^\mu \,. \label{eq:descent2c}
\end{align}

\subsection{Consistency of the deformed gauge algebra}

At antifield number 2, the descent equation \eqref{eq:descent2a} dictates that $(a_2,a_2)$ must be $\gamma$-exact modulo a total divergence. We recall here the full expression for the candidate $a_2$ that meets the consistency requirements at first order,
\begin{equation}
a_2=a_2^{\rm(EH)}+a_2^{\rm(YM)}+a_2^{({\rm PM}-g)} \,,
\end{equation}
respectively with structure constants $g_{IJK}=g_{(IJK)}$, $f_{abc}=f_{[abc]}$ and $a^I{}_{\Delta\Omega}=a^I{}_{(\Delta\Omega)}$.
Consistency of this $a_2$ with Eq.\ \eqref{eq:descent2a} yields the following set of quadratic constraints:
\begin{equation} \label{eq:quadratic constraints antifield2}
\begin{aligned}
& 0=g^I{}_{M[J} g^M{}_{K] L} \,, \quad 0=f^c{}_{a[b} f^a{}_{d e]} \,,\\
& 0=a^{[I}{}_{\Delta\Omega}a^{J]\Delta}{}_{\Theta} \,,\quad 0=a_{I\Delta[\Omega}a^I{}_{\Theta]\Sigma} \,,\quad 0=a^{(I}{}_{\Delta\Omega}a^{J)\Delta}{}_{\Theta}+\tfrac{1}{4}\,g^{KIJ}a_{K\Omega\Theta} \,. 
\end{aligned}
\end{equation}
The first among these results states that the multi-graviton algebra is associative (in addition to being commutative and symmetric as per the constraint $g_{IJK}=g_{(IJK)}$) \cite{Boulanger:2000rq}. The second result is the usual Jacobi identity of Yang-Mills theory. The second line contains new results corresponding to restrictions on the PM gravitational coupling.

\subsection{Consistency of the deformed gauge symmetries}

Next we analyze the consistency of the gauge symmetry deformation $a_1$ with the second descent equation, Eq.\ \eqref{eq:descent2b}. Our candidate $a_1$ is given by
\begin{equation}
a_1=a_1^{\rm(EH)}+a_1^{\rm(YM)}+a_1^{({\rm PM}-g)}+\bar{a}_1^{\rm(PM)}+\bar{a}_1^{(v-g)}+\bar{a}_1^{(v-{\rm PM})}+\bar{a}_1^{\rm(non-geo)} \,.
\end{equation}
We remind the reader that the last four of these terms correspond to Abelian deformations, with coefficients $c_{\Sigma\Delta\Omega}$, $d_{Iab}$, $e_{\Delta ab}$ and $b_{I\Delta\Omega}$, respectively.

Several among the obstructions (i.e.\ terms in \eqref{eq:descent2b} which are neither 
$\gamma$-exact nor $\delta$-exact modulo total derivatives) are eliminated thanks to 
the constraints derived previously, Eq.\ \eqref{eq:quadratic constraints antifield2}. 
The remaining obstructions necessitate the following constraints on the Abelian coefficients:
\begin{equation} \label{eq:quadratic constraints antifield1}
\begin{aligned}
& 0=f^{a}{}_{b(c}d^{I}{}_{d)a} \,,\quad  0=f^{a}{}_{b(c}e^{\Delta}{}_{d)a} \,,\\
& 0= d_{Iab}a^{I\Delta\Omega}-4e^{(\Delta}{}_{ac}e^{\Omega) c}{}_{b} \,,
\quad 0=e^{[\Delta}{}_{ab} e^{\Omega] ac}{} \,,\\
& 0=e^{\Delta a}{}_{(b}d^{I}{}_{c)a}+4 e_{\Omega bc}a^{I\Delta\Omega} \,, 
\quad 0=e^{\Delta a}{}_{[b}d^{I}{}_{c]a} \,,\\
& 0=d_{(I}{}^{ab}d_{J)ac}-g_{IJK}d^{Kb}{}_{c} \,,\quad 0=d_{[I}{}^{ab}d_{J]ac} \,,\\
& 0= c_{\Delta\Sigma}{}^{\Omega}c_{\Theta\Pi\Omega}
+4\,a_{I\Delta\Sigma}a^{I}{}_{\Theta\Pi} \,,\\
& 0=a_{I\Sigma[\Pi}c_{\Delta]\Omega}{}^{\Sigma} \,, \quad 
0=b_{I\Delta\Omega}b^{J\Delta}{}_{\Pi} \,.
\end{aligned}
\end{equation}
The quadratic constraints on $c_{\Sigma\Delta\Omega}$ generalize the results derived in Ref.\ \cite{Boulanger:2019zic} to include gravitational couplings; to our knowledge, the rest are new results.

\subsection{Consistency of the cubic vertices}

Considering finally the descent equation \eqref{eq:descent2c}, we find that most 
obstructions are canceled upon use of the constraints derived above, 
Eqs.\ \eqref{eq:quadratic constraints antifield2} and 
\eqref{eq:quadratic constraints antifield1}. This is a highly non-trivial consistency 
check given that a priori the antibracket $(a_1,a_0)$ contains several hundreds of terms. 
The offending terms that remain read
\beq\bal \label{eq:antifield0 obstruction}
(a_1,a_0)&\supset \left(c_{\Delta\Sigma}{}^{\Omega}c_{\Omega\Theta\Pi}+4a_{I\Delta\Sigma}a^{I}{}_{\Theta\Pi}\right)\F^{\Theta\mu}{}_{\rho\sigma}\F^{\Pi\nu\rho\sigma}\F^{\Sigma\lambda}{}_{\mu\nu}\nabla_{\lambda}\chi^{\Delta} \\
&\quad +\left(2e^{\Omega}{} _{ab}c_{\Delta\Sigma\Omega}+d^{I}{}_{ab}a_{I\Delta\Sigma} \right)F^{a\mu\rho}F^{b}{}_{\rho}{}^{\nu}\F^{\Sigma\lambda}{}_{\mu\nu}\nabla_{\lambda}\chi^{\Delta} \,,
\eal\eeq
leading to the following constraints:
\begin{equation} \label{eq:quadratic constraints antifield0}
\begin{aligned}
& 0=c_{\Delta\Sigma}{}^{\Omega}c_{\Omega\Theta\Pi}+4a_{I\Delta\Sigma}a^{I}{}_{\Theta\Pi} \,,\\
& 0=2e^{\Omega}{} _{ab}c_{(\Delta\Sigma)\Omega}+d^{I}{}_{ab}a_{I\Delta\Sigma} \,,\quad 0=e^{\Omega}{} _{ab}c_{[\Delta\Sigma]\Omega} \,.
\end{aligned}
\end{equation}
The constraints in the first line were also given previously in 
Ref.\ \cite{Boulanger:2019zic} with $a_{I\Delta\Sigma}=0$. However, in that reference, 
this result was assumed to hold rather than derived. 
Here we analyze this and the other obstruction more closely. 
The combination $\F^{\Theta\mu}{}_{\rho\sigma}\F^{\Pi\nu\rho\sigma}\F^{\Sigma\lambda}{}_{\mu\nu}\nabla_{\lambda}\chi^{\Delta}$ is clearly not $\delta$-exact since it does not vanish on the free-field equations of motion (even allowing for total derivatives). That it is also not $\gamma$-exact can be seen by considering the flat-space limit, in which case the subexpression $\F^{\Theta\mu}{}_{\rho\sigma}\F^{\Pi\nu\rho\sigma}\F^{\Delta\lambda}{}_{\mu\nu}$ would need to be a total derivative in order to get, upon partial integration, a second derivative acting on the ghost. That this is not the case can be verified directly by calculating its Euler-Lagrange derivative. The other obstruction in \eqref{eq:antifield0 obstruction} may be analyzed in the same way.

%%%%%%%%%%%%%%%%%%%%%%%%%%%%
%%%%%%%%%%%%%%%%%%%%%%%%%%%%

\section{Analysis of the results} \label{sec:analysis}

The list of quadratic constraints given in Eqs.\ \eqref{eq:quadratic constraints antifield2}, \eqref{eq:quadratic constraints antifield1} and \eqref{eq:quadratic constraints antifield0} constitute the main results of this paper. In this Section we analyze the resolution of these constraints. Our aim is not to be exhaustive but rather to understand the implications of assuming versus relaxing the condition of having sign-definite internal metrics. We will show that this assumption is inconsistent with the existence of most of the cubic vertices, the unique non-trivial exception being the case of multiple independent conformal gravity sectors coupled to Yang-Mills theory in $D=4$ dimensions.

\paragraph{Massless spin-1 and PM spin-2.} For the sake of clarity, and because it is an interesting system on its own, we consider first the couplings of massless spin-1 and PM spin-2 fields. We repeat here the pertinent constraints (omitting the usual Jacobi identity for $f_{abc}$ and noting that $a_{I\Delta\Sigma}=0$ as we ignore gravity here):
\begin{equation} \label{eq:only PM spin-1 constraints}
\begin{aligned}
& 0=c_{\Delta\Sigma}{}^{\Omega}c_{\Theta\Pi\Omega} \,, \quad 0=c_{\Delta\Sigma}{}^{\Omega}c_{\Omega\Theta\Pi} \,,\\
& 0=e^{\Delta}{}_{ac}e^{\Omega c}{}_{b} \,,\quad 0=f^{a}{}_{b(c}e^{\Delta}{}_{d)a} \,,
\end{aligned}
\end{equation}
and recall that $D=4$ for the coefficients $c_{\Sigma\Delta\Omega}$ and $e_{\Delta ab}$ to be a priori non-zero. Non-vanishing solutions to these constraints do not exist in the case of sign-definite metrics, i.e.\ when $\mathfrak{g}_{ab} = \pm \delta_{ab}$ and $\mathfrak{g}_{\Delta\Omega} = \pm \delta_{\Delta\Omega}$. The argument is the same as the one given in Ref.\ \cite{Boulanger:2019zic}: considering $\Sigma=\Theta$ and $\Delta=\Pi$ (with no summation), the contraction $c_{\Delta\Sigma}{}^{\Omega}c_{\Theta\Pi\Omega}$ becomes a sum of squares, hence $c_{\Sigma\Delta\Omega}=0$. Similarly $e_{\Delta ab}=0$. An obvious corollary is that a single PM field cannot interact via cubic vertices with a unitary Yang-Mills (or Maxwell) sector.

The conclusion is different if one allows for `healthy/ghostly' relative signs. Explicit solutions for the PM spin-2 self-coupling were given in \cite{Boulanger:2019zic}. Here we only consider the mutual vector-PM interaction. In order to have a non-vanishing $e_{\Delta ab}$ we need a non-sign-definite internal metric $\mathfrak{g}_{a b}$. The most minimal case includes one PM field and two Abelian vector fields ($f_{abc}=0$), with $\mathfrak{g}_{a b}={\rm diag}(+1,-1)$. The unique solution (modulo a trivial overall rescaling) is given by $e_{ab}:=e_{1ab}=1\;\forall\,a,b$. If we consider three vector fields (the simplest case that \textit{a priori} allows for non-zero $f_{abc}$) with $\mathfrak{g}_{a b}={\rm diag}(+1,+1,-1)$, and again a single PM field, we find a family of non-trivial solutions for $e_{ab}$; however, they all lead to a vanishing $f_{abc}$ as per the last constraint in \eqref{eq:only PM spin-1 constraints}.

To have a non-zero `ghostly Yang-Mills' coupling one needs at least four vectors. We have found the general solution in this case (still under the assumption of one PM field); it involves several free parameters and rather lengthy expressions, so for brevity we do not write here the full result and instead only give a particular solution in the model with metric $\mathfrak{g}_{a b}={\rm diag}(+1,+1,+1,-1)$:
\begin{equation}
\begin{aligned}
&\,f_{123}=1\,,\quad f_{124}=f_{134}=\frac{1}{2}\,,\quad f_{234}=\frac{1}{\sqrt{2}} \,,\quad e_{11}=e_{34}=-e_{24}=1 \,,\\
&\,e_{12}=-e_{13}=-\frac{1}{\sqrt{2}}\,,\quad e_{14}=\sqrt{2}\,,\quad e_{22}=e_{33}=-e_{23}=\frac{1}{2}\,,\quad e_{44}=2 \,.
\end{aligned}
\end{equation}

\paragraph{Massless spin-1 and massless spin-2.} We study next the mutual couplings of massless spin-2 fields mediated by vector particles. We assume sign-definite metrics $\mathfrak{g}_{ab} = \pm \delta_{ab}$ and $\mathfrak{g}_{IJ} = \pm \delta_{IJ}$. We start by recalling the theorem of Ref.\ \cite{Boulanger:2000rq} stating that the unique solution (modulo overall rescalings) of the constraints on the multi-graviton coefficients $g_{IJK}$ is given by $g_{IJK}=1$ if $I=J=K$ and $g_{IJK}=0$ otherwise.

The constraint $0=d_{(I}{}^{ab}d_{J)ac}-g_{IJK}d^{Kb}{}_{c}$ (cf.\ Eq.\ \eqref{eq:quadratic constraints antifield1}) may then be written as $\left(d_I d_J\right)_{bc}= \pm \delta_{IJ}\left(d_I\right)_{bc}$ (no sum over $I$), where $(d_I)$ is the matrix with entries $d_{Iab}$. This result shows that these matrices are $n_g$ independent projectors. As a consequence, a basis of solutions is given by $d_{Iab}=\pm \delta_{Ia}\delta_{Ib}$ (no sum over $I$). The other relevant constraints, $0=d_{[I}{}^{ab}d_{J]ac}$ and $0=f^{a}{}_{b(c}d^{I}{}_{d)a}$, are then automatically satisfied.
It follows that a massless spin-1 particle may only couple to one graviton (or to none), 
thus forbidding vector-mediated multi-graviton interactions. 
This also applies to Yang-Mills theory: if $d_{Iab}\neq0$, then the constraint 
$0=f^{a}{}_{b(c}d^{I}{}_{d)a}$ is satisfied, with non-zero $f_{abc}$, only if the `$a$' 
and `$b$' vector fields belong to the same Yang-Mills sector. Put another way, while 
two non-Abelian vectors may couple to distinct gravitons, they cannot be components of 
the same Yang-Mills multiplet, i.e.\ with a ``common'' $f_{abc}$. This precludes the 
possibility of multi-graviton interactions through loops of vector particles.

This outcome generalizes the no-go theorem of \cite{Boulanger:2000rq}, which considered 
couplings mediated by scalar particles, and is also in agreement with the results of 
Ref.\ \cite{Bizdadea:2007vh}. The latter analysis was however restricted to the Abelian 
spin-1 case; the present no-go result for Yang-Mills theory appears to be new to the 
best of our knowledge. 
Once again these results are not expected to uphold in the situation with non-sign-definite 
internal metrics. In fact, already for pure multi-gravity non-trivial couplings are known to 
exist if one allows for `ghostly' massless spin-2 fields \cite{Cutler:1986dv,Wald:1986dw}.

\paragraph{Massless spin-2 and PM spin-2.} Focusing first on the geometric
coupling between massless and PM gravitons, we have the quadratic constraints 
(cf.\ Eqs.\ \eqref{eq:quadratic constraints antifield2}, 
\eqref{eq:quadratic constraints antifield1} and \eqref{eq:quadratic constraints antifield0})
\begin{equation}
\begin{aligned}
& 0=a^{[I}{}_{\Delta\Omega}a^{J]\Delta}{}_{\Theta} \,,\quad 0=a_{I\Delta[\Omega}a^I{}_{\Theta]\Sigma} \,,\quad a^{(I}{}_{\Delta\Omega}a^{J)\Delta}{}_{\Theta}=-\frac{1}{4}g^{KIJ}a_{K\Omega\Theta} \,,\\
& 0=c_{\Delta\Sigma}{}^{\Omega}c_{\Theta\Pi\Omega}+4a_{I\Delta\Sigma}a^{I}{}_{\Theta\Pi} \,,\quad 0=a_{I\Sigma[\Pi}c_{\Delta]\Omega}{}^{\Sigma} \,,\\
& 0=c_{\Delta\Sigma}{}^{\Omega}c_{\Omega\Theta\Pi}+4a_{I\Delta\Sigma}a^{I}{}_{\Theta\Pi} \,.
\end{aligned}
\end{equation}
Assuming sign-definite kinetic metrics, these constraints may be analyzed in a similar fashion to the vector-graviton interaction of the previous paragraph, although here we need to be more cautious about the relative kinetic signs between massless and PM sectors. 
Without loss of generality, we assume the former to be 
$\mathfrak{g}_{IJ}=\delta_{IJ}$, 
and write $\mathfrak{g}_{\Delta\Sigma}=\sigma_{\rm PM}\delta_{\Delta\Sigma}$ with $\sigma_{\rm PM}=\pm1$. The constraint $a^{(I}{}_{\Delta\Omega}a^{J)\Delta}{}_{\Theta}=-\frac{1}{4}g^{KIJ}a_{K\Omega\Theta}$ is then solved by $a_{I\Delta\Omega}=-\frac{1}{4}\sigma_{\rm PM}g_I\delta_{I\Delta}\delta_{I\Omega}$ (no sum over $I$), where $g_I:= g_{III}$. The constraints involving the PM self-coupling $c_{\Delta\Sigma\Omega}$ may be manipulated to produce $0=c_{[\Delta\Sigma]}{}^{\Omega}c_{\Theta\Pi\Omega}$, which implies that $c_{[\Delta\Sigma]\Omega}=0$ if the internal metric is sign-definite. Thus we reach the conclusion that
\beq
c_{\Delta\Sigma\Omega}=c_{(\Delta\Sigma\Omega)} \,.
\eeq
From these results it can be demonstrated that one may choose a basis in which $c_{\Delta\Sigma\Omega}=0$ unless $\Delta=\Sigma=\Omega$. Indeed, from the constraints $0=c_{\Delta(\Sigma}{}^{\Omega}c_{\Theta)\Pi\Omega}+4a_{I\Delta(\Sigma}a^{I}{}_{\Theta)\Pi}$ and $0=c_{\Delta[\Sigma}{}^{\Omega}c_{|\Omega|\Theta]\Pi}$ we infer that $c_{\Delta\Delta}{}^{\Omega}c_{\Sigma\Sigma\Omega}=0$ (no sum over $\Delta,\Sigma$) if $\Delta\neq\Sigma$. Thus $\{c_{(\Delta)}\}_{\Delta=1}^{n_{\rm PM}}$, where $c_{(\Delta)}$ is the vector with components $c_{\Delta\Delta}{}^{\Omega}$, is an orthogonal set, and we may choose a basis with $(c_{(\Delta)})^{\Omega}\propto \delta^{\Omega}_{\Delta}$. Then, using this result and the constraints, we find (up to a sign)
\beq
c_{\Delta\Delta\Delta}=\frac{\sqrt{-\sigma_{\rm PM}}}{2}g_{\Delta} \,,
\eeq
where $g_{\Delta}:=g_I\delta_{I\Delta}$ is non-zero only if the `$\Delta$' PM field couples to one of the massless gravitons. Notice furthermore that we are forced to consider a `ghostly' PM sector with $\sigma_{\rm PM}=-1$ in order to have a real solution. Finally, we can now use this result once again in the original constraint to infer that $c_{\Delta\Sigma\Omega}=0$ when $\Delta\neq\Sigma\neq\Omega$. Note that implicit in this analysis is the assumption that $D=4$. If $D\neq4$ then $c_{\Delta\Sigma\Omega}=0$ from the start, and it is then easy to prove from the above constraints that $a_{I\Delta\Sigma}=0$ in this case.

In conclusion, a PM spin-2 field may only interact with at most one graviton, and only in four dimensions, at least if one supposes the existence of a cubic vertex as dictated by the minimal coupling prescription, while mutual interactions among different PM fields or among different massless spin-2 fields are excluded. The single PM-graviton system is consistent with the expectations spelled out in the Introduction, since we know that conformal gravity precisely includes a massless and a PM spin-2 fields, which must have opposite kinetic signs. The obstruction to mutual couplings between different conformal gravity sectors is also in agreement with the general results of \cite{Boulanger:2001he}.

The non-geometric coupling is, on the other hand, fully obstructed in the situation with 
sign-definite metrics, since the constraint $0=b_{I\Delta\Omega}b^{J\Delta}{}_{\Pi}$ then 
implies $b_{I\Delta\Omega}=0$, as we have explained. An immediate corollary is that a 
single PM field cannot interact with gravity through this Abelian vertex. 
The failure of the non-geometric coupling may be traced back to the absence of a mixed 
`$g-b$' term in the quadratic constraint (as present in the case of the geometric couplings), 
which is due to the absence of 
a non-Abelian (i.e., non-trivial $a_2$) deformation for the non-geometric coupling. 
We find it interesting, in this respect, that non-Abelian deformations are in some sense 
less constrained than Abelian ones.

Dropping the hypothesis of sign-definite metrics again changes this no-go result. The simplest such model requires one massless graviton and two relatively `ghostly' PM spin-2 particles with $\mathfrak{g}_{\Delta\Omega}={\rm diag}(+1,-1)$, leading to the non-trivial solution $b_{\Delta\Omega}:=b_{1\Delta\Omega}=1\; \forall\Delta,\Omega$ (modulo an overall rescaling). More complex models with more than two fields may be similarly studied, with solutions analogous to those given in Ref.\ \cite{Boulanger:2019zic}.

\paragraph{General case with sign-definite kinetic terms.} Finally we consider the general case with massless spin-1, massless spin-2 and PM spin-2 fields, focusing exclusively on the case of sign-definite internal metrics, although we allow for relative signs between different particle types. As we have seen, this assumption forbids the PM non-geometric coupling ($b_{I\Delta\Omega}=0$). We write the internal metrics as $\mathfrak{g}_{IJ}=\delta_{IJ}$, $\mathfrak{g}_{\Delta\Sigma}=\sigma_{\rm PM}\delta_{\Delta\Sigma}$, $\mathfrak{g}_{ab}=\sigma_v \delta_{ab}$, where $\sigma_{\rm PM}$ and $\sigma_v$ give the relative kinetic signs of the PM and vector sectors.

Consider first, for the sake of clarity, the situation with only one field of each type, so we omit all indices in the structure coefficients (assuming the convention that all indices have been lowered with the internal metrics). Normalizing $g=1$, we have from \eqref{eq:quadratic constraints antifield2} $a=-\sigma_{\rm PM}/4$ (we ignore trivial solutions, here $a=0$). The PM self-coupling may be analyzed as in the previous paragraph, with the result $c=\sqrt{-\sigma_{\rm PM}}/2$, so again $\sigma_{\rm PM}=-1$ for the solution to be real (and recall that $D=4$ is also necessary). We are left to consider the mixed constraints involving $d$ and $e$. There are four constraints in total (three in \eqref{eq:quadratic constraints antifield1} and one in \eqref{eq:quadratic constraints antifield0}), i.e.\ it is an over-determined system, yet a (unique) solution exists: $d=\sigma_v$, $e=\sigma_v/4$. Notice that the sign $\sigma_v$ remains undetermined, i.e.\ both `healthy/ghostly' cases for the vector field are allowed. This outcome agrees with the expectation inferred from the existence of conformal gravity coupled to a Maxwell field, which maintains conformal invariance in four dimensions. The strength of this result lies in having established the uniqueness of the solution.

The generalization of this analysis to multiple fields is straightforward upon use of the previous results in this section. We first use the results of the massless-PM system to infer that $a_{I\Delta\Omega}=\frac{1}{4}g_I\delta_{I\Delta}\delta_{I\Omega}$ and $c_{\Delta\Sigma\Omega}=\frac{1}{2}g_I\delta_{I\Delta}\delta_{I\Sigma}\delta_{I\Omega}$, with the requirement that $\sigma_{\rm PM}=-1$ (and we choose $\sigma_v=1$ for concreteness), implying in particular that massless-PM spin-2 fields may only couple in independent pairs (or else remain uncoupled). The massless spin-1 gravitational coupling is also studied in the same way as above, i.e.\ $d_{Iab}=g_I\delta_{Ia}\delta_{Ib}$, and identical conclusions follow. The remaining non-trivial constraints involving $e_{\Delta ab}$ are
\begin{equation}
\begin{aligned}
& 0=f^{a}{}_{b(c}e^{\Delta}{}_{d)a} \,,\quad 0= d_{Iab}a^{I\Delta\Omega}-4e^{(\Delta}{}_{ac}e^{\Omega) c}{}_{b} \,,\quad 0=e^{[\Delta}{}_{ab} e^{\Omega] ac}{} \,,\\
& 0=e^{\Delta a}{}_{(b}d^{I}{}_{c)a}+4 e_{\Omega bc}a^{I\Delta\Omega} \,,\quad 0=2e^{\Omega}{} _{ab}c_{(\Delta\Sigma)\Omega}+d^{I}{}_{ab}a_{I\Delta\Sigma} \,.
\end{aligned}
\end{equation}
Consider first the second-to-last of these equations, and fix the free index `$I$' here to correspond to a graviton which does \textit{not} couple to a PM field. It then follows that $e_{\Delta ab}=0$, i.e.\ the massless vectors cannot interact with an isolated PM field, in agreement with our previous findings. If on the other hand we have a non-trivial massless-PM spin-2 pair with $\Delta=I$ (in a suitable basis), then this constraint allows one to show that $e_{\Delta ab}\propto \delta_{I\Delta}\delta_{Ia}\delta_{Ib}$. The proportionality constant is fixed by the other constraints: $e_{\Delta ab}=\frac{1}{4}g_I\delta_{I\Delta}\delta_{Ia}\delta_{Ib}$. This establishes that distinct conformal gravity sectors cannot mutually couple through vector particles. This also applies to interactions mediated by loops of spin-1 fields belonging to the same Yang-Mills multiplet as per the constraint $0=f^{a}{}_{b(c}e^{\Delta}{}_{d)a}$, using the same reasoning we used previously in the analysis of the vector-graviton system.

%%%%%%%%%%%%%%%%%%%%%%%%%%%%
\section{Conclusions}\label{sec:conclusions}

In this paper, we provided a complete classification of the 
consistent first-order deformations of the free theory 
describing an arbitrary collection of massless spin-1, 
massless spin-2 and partially-massless (PM) spin-2 fields 
in rigid $D$-dimensional (A)dS space. As our sole assumptions we 
requested the vertices to be parity-even, to contain no more 
than two derivatives and to respect the isometries of the (A)dS 
background of the free theory.

As far as interactions among massless and PM spin-2 fields are concerned, 
our results confirm the classification of Ref.\ \cite{Joung:2019wwf} 
with what it called the geometric, non-Abelian coupling 
\eqref{geom_vertex1}-\eqref{geom_vertex2}, as well as the 
non-geometric Abelian coupling \eqref{non-geom_vertex1}-\eqref{non-geom_vertex2}.
Under our assumptions on the number of derivatives in the vertices, 
we also find a unique candidate vertex mixing massless spin-1 
and PM spin-2 particles, see \eqref{Chaplin-Manton-like}.
This vertex is of the Chapline-Manton type, i.e., 
it is Abelian yet induces a non-linear gauge transformation of 
the spin-1 fields. It only exists in $D=4$ dimensions and 
mimics the minimal gravitational coupling of Maxwell's fields, now 
for PM spin-2 fields instead of massless spin-2 fields. 

In Sec.\ \ref{sec:secondorder} 
we analyzed all the consistency conditions of the candidate deformations 
at second order in perturbation, thereby producing the complete set of quadratic 
constraints on the structure constants that appear at first order in deformation, 
see \eqref{eq:quadratic constraints antifield2}, \eqref{eq:quadratic constraints antifield1}
and \eqref{eq:quadratic constraints antifield0}.
We considered, in Sec.\ \ref{sec:analysis}, the most general solution of these constraints 
under the assumption that each field sector contains no relative healthy/ghostly 
signs in the kinetic terms, although distinct sectors may do so. 
The solution is given by multiple, independent copies of $D=4$ conformal gravity 
minimally coupled with a Yang-Mills (or possibly Abelian) spin-1 sector.
Our findings allow us to rule out the non-geometric vertex, 
at least under the aforementioned assumptions. 
We could also generalize the no-go theorem of Ref.\ \cite{Boulanger:2000rq}
by showing that distinct massless graviton species cannot mutually interact 
through the exchange of massless spin-1 (Abelian or Yang-Mills) or PM spin-2 particles.

Finally, in Sec.\ \ref{sec:analysis} we also exhibited some solutions to the quadratic constraints 
\eqref{eq:quadratic constraints antifield2}, \eqref{eq:quadratic constraints antifield1}
and \eqref{eq:quadratic constraints antifield0},
in the general set-up with non-sign-definite internal metrics.  
We were able to exhibit particular solutions for which all the candidate 
vertices remain consistent. We speculate that, similarly to what was done 
for the pure PM spin-2 case studied in \cite{Boulanger:2019zic}, these solutions 
with non-sign-definite internal metrics give rise to full theories, complete 
at the cubic order, consistent as far as the preservation of number of degrees of 
freedom is concerned. We hope to be able to report on this point in the near future.

%%%%%%%%%%%%%%%%%%%%%%%%%%%%

\acknowledgments

The work of N.B.\ was partially supported by the F.R.S.-FNRS PDR Grant 
No.\ T.0022.19 ``Fundamental issues in extended gravity'', Belgium. 
During the accomplishment of the work, L.T.\ was a Research Fellow (ASP) of the Fund 
for Scientific Research-FNRS, Belgium.
S.G.S.\ and S.S.P.\ acknowledge support from the NSFC Research Fund for International 
Scientists (Grant No.\ 12250410250).

%%%%%%%%%%%%%%%%%%%%%%%%%%%%
%%%%%%%%%%%%%%%%%%%%%%%%%%%%

\appendix

\section{Partially massless spin-2 field equations} 
\label{sec:appendix}

The classification of non-trivial cubic vertices requires knowledge of the cohomology of the differential $\delta$. It proves useful to this end to know all the consequences of the free-field equations of motion for our system, since trivial cubic vertices are precisely ones that vanish on these equations. In this Appendix we analyze the free PM spin-2 theory, the cases of massless spin-1 and spin-2 being of course very well known.

The equation of motion that derives from the free PM action is
\beq
{\cal E}^{\mu\nu}:=\frac{1}{\sqrt{-g}}\,\frac{\delta S_0}{\delta k_{\mu\nu}}=\nabla_{\rho}\F^{\rho(\mu\nu)}-g^{\mu\nu}\nabla_{\rho}\F^{\rho}+\nabla^{(\mu}\F^{\nu)} \,.
\eeq
We omit the color index which obviously plays no role here. The Noether identity that follows from the PM gauge symmetry is
\beq
\nabla_{\mu}\nabla_{\nu}{\cal E}^{\mu\nu}-\frac{\sigma}{L^2}\,{\cal E}=0 \,,
\eeq
with ${\cal E}:=g_{\mu\nu}{\cal E}^{\mu\nu}$.

The PM field strength and its derivatives satisfy several identities in terms of ${\cal E}^{\mu\nu}$:
\beq \label{eq:onshell1}
\F^{\mu}=\nabla^{\mu}k-\nabla_{\nu}k^{\mu\nu}=-\frac{\sigma L^2}{D-2}\,\nabla_{\nu}{\cal E}^{\mu\nu}\,,\qquad \nabla_{\mu}\F^{\mu}=-\frac{1}{D-2}\,{\cal E}\,,
\eeq
\beq \label{eq:onshell2}
\nabla_{\rho}\F^{\mu\nu\rho}=\frac{2\sigma L^2}{D-2}\,\nabla^{[\mu}\nabla_{\rho}{\cal E}^{\nu]\rho}\,,\qquad \nabla_{\rho}\F^{\rho\mu\nu}={\cal E}^{\mu\nu}-\frac{g^{\mu\nu}}{D-2}\,{\cal E}+\frac{\sigma L^2}{D-2}\,\nabla^{\nu}\nabla_{\rho}{\cal E}^{\mu\rho}\,,
\eeq
\beq
\Box \F_{\mu\nu\rho}+\frac{(2D-3)\sigma}{L^2}\,\F_{\mu\nu\rho}=2\nabla_{[\mu}{\cal E}_{\nu]\rho}+\frac{2}{D-2}\,g_{\rho[\mu}\nabla_{\nu]}{\cal E}\,.
\eeq
It follows in particular that the trace and divergences of $\F_{\mu\nu\rho}$ all vanish on the equations of motion.

Another identity related to \eqref{eq:onshell2} is
\beq
\Box k_{\mu\nu}-\nabla_{\mu}\nabla_{\nu}k-\frac{\sigma}{L^2}(g_{\mu\nu}k-D k_{\mu\nu})={\cal E}_{\mu\nu}-\frac{g_{\mu\nu}}{D-2}\,{\cal E}+\frac{2\sigma L^2}{D-2}\,\nabla_{(\mu}\nabla^{\rho}{\cal E}_{\nu)\rho}\,.
\eeq
On the equations of motion, and considering specifically the gauge $k=0$ (which from \eqref{eq:onshell1}, incidentally, implies also $\nabla_{\nu}k^{\mu\nu}=0$), the previous equation reduces to the standard wave form (see e.g.\ \cite{Deser:2001us}):
\beq
\left(\Box+\frac{\sigma D}{L^2}\right)k_{\mu\nu}=0\qquad\qquad \left(\mathcal{E}^{\mu\nu}=0\,,\; k=0\right) \,.
\eeq

%%%%%%%%%%%%%%%%%%%%%%%%%%
%%%%%%%%%%%%%%%%%%%%%%%%%%

\bibliographystyle{JHEP}
\bibliography{PM-YM-gravity_refs}

\end{document}